\documentclass[preprint, tighten, twocolumn]{aastex63}

\usepackage[clockwise,figuresright]{rotating}

\usepackage{amsfonts,amsmath,amssymb}

\usepackage{lineno}
\UseRawInputEncoding

\newcommand{\W}{\mbox{$\mathcal{W}$}}

\received{}
\revised{}
\accepted{}

\submitjournal{ApJ}

\shorttitle{MHD Winds Driven by Line Force}
\shortauthors{Yang et al.}

\begin{document}

\title{Magnetohydrodynamic Winds Driven by the Line Force from the Standard Thin Disk around Supermassive Black Holes. I. The Case of Weak Magnetic Field}

\correspondingauthor{Xiao-Hong Yang}
\email{yangxh@cqu.edu.cn}

\author[0000-0002-2419-9590]{Xiao-Hong Yang}
\author{Kamarjan Ablimit}
and
\author[0000-0003-3706-5652]{Qi-Xiu Li}
\affiliation{Department of Physics, Chongqing University, Chongqing 400044, People's Republic of China; yangxh@cqu.edu.cn}

\begin{abstract}
Absorption lines with high blue-shifted velocities are frequently found in the ultraviolet (UV) and X-ray spectra of luminous active galactic nuclei (AGNs). This implies that high-velocity winds/outflows are common in AGNs. In order to study the formation of high-velocity winds, especially ultrafast outflows (UFOs), we perform two-dimensional magnetohydrodynamic (MHD) simulations. Initially, a magnetic field is set to be weaker than the gas pressure at the disk surface. In our simulations, line force operates on the region like filaments because the X-ray radiation from corona is shielded by dense gas in the inner region at some angle. The location of filaments changes with time and then the line-driven winds are exposed to X-ray and become highly ionized. The line force at the UV bands does not directly drive the highly ionized winds. In the sense of time average, the properties of high-velocity winds meet the formation condition of UFOs. Compared with line force, the function of magnetic field is negligible in directly driving winds. In the MHD model, the region around the rotational axis becomes magnetic-pressure dominated, which prevents gases from spreading to higher latitudes and then enhances the gas column density at middle and low latitudes (20$^{\rm o}$--70$^{\rm o}$). Higher column density is helpful to shield X-ray photons, which causes the line force to be more effective in the MHD model than in the hydrodynamic model. Higher-velocity winds with a broader opening angle are produced in the MHD model.
\end{abstract}


\section{Introduction} \label{sec:intro}
Observations frequently find the blue-shifted absorption lines in the ultraviolet (UV) and X-ray spectra of luminous active galactic nuclei (AGNs). The blue-shifted absorption lines imply that winds/outflows are common in AGNs. According to line widths, the absorption lines have been generally classified as broad absorption lines (BALs) with the typical widths of several thousands kilometers per second, narrow absorption lines (NALs) with typical widths of a few hundred kilometers per second, and the so-called \textit{min-BALs} with intermediate widths (Churchill 1999; Narayanan et al. 2004; Chartas et al. 2009a; Rodr\'{i}guez Hidalgo et al. 2011). The widths of intrinsic absorption lines may depend on the properties of absorbers, or the line of sight through the same absorbers (Ganguly et al. 2001).

At the UV band, Hubble Space Telescope (HST) spectra indicate that approximately half of the Seyfert 1 galaxies have UV NALs from C \footnotesize{\rm IV}\normalsize{ } and N \footnotesize{\rm V}\normalsize{ }, which are all blue-shifted by zero to -1500 km s$^{-1}$  (e.g., Crenshaw 1997). Blueshifted lines are believed to be formed in fast outflows. Highly blue-shifted lines are also found. Hamann et al. (1997) reported three quasi-stellar objects (QSOs), that have intrinsic UV NAL absorbers with line-of-sight velocities from $\sim$1500 km s$^{-1}$ to $\sim$51000 km s$^{-1}$. From 37 optically bright quasars at $z$=2--4, Misawa et al. (2007) found that 10--17\% of C \footnotesize{\rm IV}\normalsize{} NALs are blue-shifted at 5000--70,000 km s$^{-1}$. Some quasars also have highly blue-shifted mini-BALs and BALs (Rodr\'{i}guez Hidalgo et al. 2011; Narayanan et al. 2004; Hamann \& Sabra 2004; Hamann et al. 2018; Xu et al. 2018). For example, a UV HST spectrum of PDS 456 contains one well-measured BAL. Hamann et al. (2018) point out that the UV BAL is probably generated by the C \footnotesize{\rm IV}\normalsize{} absorber at 0.3 $c$, where $c$ is the speed of light.

At the soft X-ray band ($<$ 3 keV), some type 1 AGNs exhibit signatures of an intrinsic ionized absorber and the absorbers are outflowing at 10--1000 km s$^{-1}$ velocities (McKernan et al. 2007). At the hard X-ray band (5--10 keV), based on XMM-Newton and Suzakue observations, blueshifted Fe XXV and Fe XXVI absorption lines are ubiquitously detected on AGN spectra (Tombesi et al. 2010; 2012; Gofford et al. 2013; 2015). The Fe XXV and Fe XXVI lines are shifted at $\sim$0.03--0.3 $c$, with a mean value of $\sim$0.14$c$, and they are generated in the interval of $10^2$--$10^4$ Schwarzschild radius ($r_{\rm s}$) from the central black hole (BH, Tombesi et al. 2012). This implies that a highly ionized absorber moves fast outward from its nucleus for $\gtrsim$40\% of AGNs. For distinguishing such outflows from fast outflows, UFOs are used to define highly ionized absorbers with outflow velocities higher than 10$^4$ Km s$^{-1}$ (Tombesi et al. 2010).

Fast outflows and UFOs may be associated with the winds produced from a standard thin disk around the supermassive BH (Proga et al. 2000; Proga \& Kallman 2004; Nomura \& Ohsuga 2017). Thermal driving, radiation driving, and magnetic driving are believed to be three important mechanisms that drive winds/outflows from the standard thin disk.

Thermal driving is a useful mechanism to understand the winds observed in some low-mass X-ray binaries (Higginbottom et al. 2017; 2018). The gas above the accretion disk is heated to the Compton temperature by X-ray photons from the corona and/or the inner disk; the gas will expand and escape, and then thermally driven winds will form. Such winds are produced outside Compton radius ($>\sim10^5$ $r_{\rm s}$). The Compton radius is used to define the location where the local isothermal sound speed at the Compton temperature equals the local escape velocity (Begelman et al. 1983). In low luminous AGNs, thermally driven winds are also expected to form at fairly large radii (Yuan \& Li 2011; Bu \& Yang 2018).

Radiation driving is a promising mechanism to understand fast outflows and UFOs. The luminous thin disk irradiates the gas above the disk, photons are scattered or absorbed by the gas, and then the radiation force is exerted on the gas. The radiation force may drive the gas to overcome the effective barrier (including the effect of gravity and centrifugal force) and then winds are produced. When the gas is highly ionized, Compton scattering is the main factor in producing the radiation force. The Compton-scattering force may be an important force to drive the winds when the effective barrier of the gas is small (e.g. Icke 1980; Tajima \& Fukue 1996; 1998; Cao 2014; Yang et al. 2018; 2019). When the gas is weakly ionized, the line absorption of UV photons is the main factor in producing the radiation force (Castor et al. 1975). Such force is called the line force. In luminous AGNs, such as quasars, the standard thin disk can emit a great number of UV photons and so the line force becomes important in driving the winds because the winds shield themselves from X-ray radiation emitted by the central corona (Murray \& Chiang 1995). A dynamical model of driving AGN winds by the line force was presented by Murray et al. (1995) and Proga et al. (2000) firstly implemented numerical simulations of AGN winds driven by the line force.

Magnetic driving is an important mechanism to explain jets or/and winds in many astrophysical environments. The magneto-centrifugally driven mechanism is suggested by Blandford \& Payne (1982). In the model, a large-scale magnetic field is required and a poloidal component ($B_{\rm p}$) is at least comparable to the toroidal component ($B_{\phi}$), i.e. $|B_{\phi}/B_{\rm p}|\lesssim1$. When the inclination angle of the field lines is less than $60^{o}$ with respect to the disk surface, the magneto-centrifugal force can drive winds from the disk. When the field lines are steep ($>60^{o}$) winds cannot be directly launched from the cold disk by the magneto-centrifugal force. Winds can be also driven by the magnetic pressure when the toroidal magnetic field is much stronger than the poloidal component, i.e. $|B_{\phi}/B_{\rm p}|\gg1$. Fukumura et al. (2018) have modeled the Fe K UFO properties in PDS 456 and pointed out that UFOs can only be driven by magnetic field.

The above three mechanisms may not operate alone in driving winds from luminous AGNs. A hybrid magnetocentrifugal and radiatively driven wind model has been proposed in the previous analytical works (Everett 2005; Cao 2014). In the previous simulation works, Proga (2003) simulated the outflows driven by line force and magnetic field. However, in Proga's work, the gas is assumed to be isothermal, and the adopted model parameters are available for young stellar objects. For AGNs, the gas above the thin disk is not isothermal. The gas becomes hot due to X-ray irradiation and/or cool due to radiation cooling. The gas temperature significantly changes with space and time. The magnitude of the line force depends on the degree of ionization and the temperature of the gas. For example, when the gas temperature is higher than $10^5$ K and the ionization parameter is higher than 10$^2$ erg s$^{-1}$ cm, the line force can be neglected. Therefore, when the heating and cooling of gas is considered, Proga's results (2003) cannot be directly used to predict the properties of AGN winds by applying the dimensionless form of hydrodynamic equations. When including the radiation heating and cooling of gas, Proga et al. (2000), Proga \& Kallman (2004), and Nomura \& Ohsuga (2017) have implemented simulations of the line-force-driven hydrodynamic (HD) winds in AGNs. However, their simulations do not include magnetic field. Waters \& Proga (2018) simulated thermally-driven MHD winds in X-ray binaries. Thermally-driven MHD winds are often generated outside the Compton radius ($\sim10^5 r_{\rm s}$) rather than around the thin disk. For AGNs, the line-force-driven MHD winds have not been studied using numerical simulations. We expect to further understand the role of magnetic field and line force in driving AGN winds. respectively and the properties of the AGN winds driven both by magnetic field and line force. Therefore, it is worth studying the line-force-driven MHD winds in AGNs using numerical simulations. In this paper, we study the MHD winds driven by radiation force (especially line force) from the thin disk around supermassive BHs, which is available for AGNs. Because we focus on the winds generated within 1500 $r_{\rm s}$, we do not consider the thermally driven winds.

The strength of magnetic field could influence the properties of winds. When the initial magnetic field is set to be too strong, simulating becomes very time-consuming or crashes due to the time step becoming too short. Therefore, this work firstly studies the case of weak magnetic field. The case of strong magnetic field needs to be studied in the next paper.

This paper is organized as follows: Section 2 introduces our model and method; Section 3 is devoted to the analysis of the simulations, and Section 4 summarizes and discusses our results.

\section{Model and Method} \label{sec:Model}
For luminous AGNs, such as quasars, a geometrically thin and optically thick disk is assumed to be around the suppermassive BH at the centre (Shakura \& Sunyaev 1973). The hot corona is suggested to be highly compact and within 10 $r_{\text{s}}$ (Reis \& Miller 2013; Uttley et al. 2014). The hot corona can irradiate the thin disk, so that the local isotropic intensity $I_{_\text{D}}(r_{_{\text{D}}})$ on the disk surface is written as
\begin{eqnarray}
\nonumber I_{_\text{D}}(r_{_\text{D}})=\frac{\varepsilon L_{\text{Edd}}}{12\pi^2 r^2_{\text{s}}}(\frac{27r_{\text{s}}^3}{r^3_{_\text{D}}}[1-(\frac{3r_{\text{s}}}{r_{_\text{D}}})^{1/2}] \\
+\frac{f_{X}}{3\pi}\{\text{arcsin}(\frac{3r_{\text{s}}}{r_{_\text{D}}})-\frac{3r_{\text{s}}}{r_{_\text{D}}}[1-(\frac{3r_{\text{s}}}{r_{_\text{D}}})^2]^{1/2}\}),
\end{eqnarray}
where $f_{\rm X}$ is the ratio of the X-ray luminosity of the corona to the disk luminosity, $r_{_\text{D}}$ is the radial position on the disk surface, $L_{\text{Edd}}$ is the Eddington luminosity, and $\varepsilon$ is the Eddington ratio of the disk luminosity (Proga et al. 1998).

Gas above the thin disk is irradiated by the radiation generated in the disk surface and hot corona. To simulate the irradiated gas, our computation region is located above the disk surface. Our simulations are implemented in a spherical coordinate ($r$,$\theta$, and $\phi$). The $\theta=\pi/2$ plane is located at the disk surface. The scale height ($H_0$) of the disk is the distance from the $\theta=\pi/2$ plane to the equatorial plane. For the classical standard thin disk, the scale height of the thin disk is not constant with the change in radius (Shakura \& Sunyaev 1973). In the inner region of the thin disk, the total pressure is dominated by radiation pressure, while the outer region is dominated by gas pressure (Shakura \& Sunyaev 1973). The change in the scale height of the disk with radius in the inner region is not more significant than that in the outer region. For example, in the radiation-dominated region, the disk scale height at 200 r$_{\rm s}$ is 30 percent higher than that at 30 r$_{\rm s}$. For simplicity of calculation, we assume the disk scale height to be constant. The scale height of the thin disk is about 3.1$\varepsilon r_{\text{s}}$ at $r=30r_{\text{s}}$. We set $H_0=3.1\varepsilon r_{\text{s}}$. This value is also adopted by Nomura \& Ohsuga (2017). We axis-symmetrically solve the following MHD equations:
\begin{equation}
\frac{d\rho}{dt}+\rho\nabla\cdot {\bf v}=0,
\label{cont}
\end{equation}

\begin{equation}
\rho\frac{d {\bf v}}{dt}=-\nabla P-\rho\nabla
\psi+\frac{1}{4 \pi}(\nabla \times \bf B)\times {\bf B}+\rho {\bf F}^{\rm rad},
\label{monentum}
\end{equation}

\begin{equation}
\rho\frac{d(e/\rho)}{dt}=-P\nabla\cdot {\bf v}+\rho \dot{E},
\label{energyequation}
\end{equation}

\begin{equation}
\frac{\partial {\bf B}}{\partial t}=\nabla\times({\bf v}\times{\bf B}).
\label{induction_equation}
\end{equation}
Here, $d/dt(\equiv \partial / \partial t+ \mathbf{v} \cdot \nabla)$ denotes the Lagrangian derivative and $\rho$ , $P$, $\mathbf{v}$, $\psi$ , and $e$ are density, pressure, velocity, gravitational potential, and internal energy, respectively, and $\bf{B}$ is the magnetic field. We employ an equation of state of ideal gas $P=(\gamma -1)e$ and set the adiabatic index $\gamma =5/3$. We also apply the pseudo-Newtonian potential, $\psi=-GM_{\rm BH}/(r_{\text{c}}-r_{\rm s})$  (Paczy\'{n}sky \& Wiita 1980), where $M_{\rm BH}$ and $G$ are the BH mass and the gravitational constant, respectively, and $r_\text{c}=\sqrt{r^2+H^2_0-2rH_0 cos(\pi-\theta)}$ is the distance from a gas source to the centre of the BH. $\mathbf{F}^{\text{rad}}$ is the radiation pressure exerted on per unit of mass due to Compton-scattering force and line force. $\rho\dot{E}$ is the net heating rate per unit of mass. For the brevity of this paper, the calculation of the radiation force is referred to Proga et al. (1998), who have introduced the details of calculations in their appendix. The net heating rate is also referred to in Equations (18)--(21) in Proga et al. (2000).

\begin{table*}
\begin{center}

\caption{Summary of models}
%%Please Capitalize the First Letter of Each Notional Word in table's caption

\begin{tabular}{ccccccccccc}
\hline\noalign{\smallskip} \hline\noalign{\smallskip}

Run & $\rho_{\rm d}$ (g.cm$^{-3}$

) & $\beta_{0}$ & $\alpha_{0}$   & $\varepsilon$  & $f_{\rm X}$ & grids &$\dot{M}_{\rm w}$ ($M_{\odot}$/yr) & $P_{\text{w}}$ (g.cm.s$^{-2}$)& $P_{\text{k,w}}$ (erg.s$^{-1}$) & $P_{\text{th,w}}$ (erg.s$^{-1}$) \\

(1) & (2)      & (3)      &  (4)  & (5)   & (6) & (7)  & (8) & (9) & (10) & (11) \\
\hline\noalign{\smallskip}
RHDW0.3  & 10$^{-12}$ & $\infty$ &/ & 0.3  & 8.9$\times10^{-2}$& 144$\times$160 &0.15  & 2.02$\times10^{34}$ &2.34$\times10^{43}$ &1.45$\times10^{41}$ \\

RHDW0.6  & 10$^{-12}$ & $\infty$ &/ & 0.6  & 6.0$\times10^{-2}$& 144$\times$160 &0.19  &1.86$\times10^{34}$ &1.68$\times10^{43}$ &8.81$\times10^{40}$  \\

RMHDW0.3 & 10$^{-12}$ & $100$ &5.0 &0.3 & 8.9$\times10^{-2}$ & 144$\times$160 & 0.18   &2.70$\times$10$^{34}$ & 3.62$\times$10$^{43}$ & 1.97$\times$10$^{41}$\\
RMHDW0.6 & 10$^{-12}$ & $100$ &5.0 &0.6 & 6.0$\times10^{-2}$ & 144$\times$160 & 0.25   &3.96$\times$10$^{34}$ & 5.95$\times$10$^{43}$ & 3.21$\times$10$^{41}$ \\
RMHDW0.6a & 10$^{-12}$ & $100$ &5.0 &0.6 & 1.2$\times10^{-1}$ & 144$\times$160 & 0.24   &3.07$\times$10$^{34}$ & 3.68$\times$10$^{43}$ & 2.01$\times$10$^{41}$ \\
RMHDW0.6h & 10$^{-12}$ & $100$ &5.0 &0.6 & 6.0$\times10^{-2}$ & 200$\times$192 & 0.26   &5.46$\times$10$^{34}$ &1.14$\times$10$^{44}$ & 7.48$\times$10$^{41}$ \\

\hline\noalign{\smallskip} \hline\noalign{\smallskip}
\end{tabular}
\end{center}

\begin{list}{}
\item\scriptsize{Column 1: model names; Column 2: the gas density at the disk surface; Columns 3 and 4: values of $\beta_0$ and $\alpha_0$, which determine the strength and inclination of the initial magnetic field, respectively; Column 5: the Eddington ratio ($\varepsilon=L_{\rm D}/L_{\rm Edd}$) of the disk luminosity; Column 6: the ratio of the X-ray luminosity to the disk luminosity; Column 7: the number of grids; Column 8: the time-averaged values of the mass outflow rate; Columns 9--11: the time-averaged values of the momentum flux, kinetic energy flux, and thermal energy flux, respectively, which are carried out by the outflows at the outer boundary. }
\end{list}
\label{tab1e_1}
\end{table*}

Based on the Sobolev approximation, the force multiplier ($\mathcal{M}$) of the line force is a function of the ionization parameter ($\xi$) and the local optical depth parameter (Rybicki \& Hummer 1978). For more details of $\mathcal{M}$, readers are referred to Equations (11)--(16) in Proga et al. (2000), where the force multiplier also depends formally on the thermal speed ($v_{\rm th}$) of gas. The thermal speed is set to 20 km s$^{-1}$, which corresponds the gas temperature of $2.5\times10^4$ K (Stevens \& Kallman 1990; Proga 2007). When the disk temperature is too low, the emitted photons cannot effectively contribute to the line force. The UV-emitting region is roughly defined as the high-temperature ($>$3000 K) part of the disk (Nomura \& Ohsuga 2017). The attenuation of the disk emissions is set to be $0.4 \text{ g}^{-1}\text{cm}^2$. The attenuation of X-ray photons depends on the ionization parameter ($\xi$) of the gas. $\xi$ is defined as $\xi=4\pi F_{_\text{X}}/n$, where $n$ is the number density of the gas ($n=\rho/\mu m_{\text{p}}$, where $\mu$ and $m_{\text{p}}$ are the mean molecular weight and the proton mass, respectively). Following Proga et al. (2000), the X-ray attenuation is set to be $0.4 \text{ g}^{-1}\text{cm}^2$ for $\xi\geq10^5$ erg s$^{-1}$ cm and $40 \text{ g}^{-1}\text{cm}^2$ for $\xi<10^5$ erg s$^{-1}$ cm and $\mu$ is set to be 1.0.

In previous works, the ratio ($f_{\text{X}}$) of the X-ray luminosity to the disk luminosity is assumed to be constant with luminosity and set to be $f_{\text{X}}=0.1$ (Proga \& Kallman 2004; Nomura \& Ohsuga 2017). However, it is proposed that an observed nonlinear relation between the X-ray (2 keV) and UV (2500 {\AA}) emissions exists in luminous AGNs (Lusso \& Risaliti 2016). This relation implies that the $f_{\text{x}}$ value is less in optically bright AGNs than in optical faint AGNs. X-ray radiation is important for the ionization of gas. Therefore, compared with the previous works, we further consider the relation of the X-ray luminosity on the UV luminosity. For simplicity, we assume that the electromagnetic emissions are produced by an extended corona and a standard thin disk. The extended corona is assumed to be spherical. The spherical corona emits only X-ray photons and the inner part of the thin disk can emit UV photons. We use the relation of X-ray and UV emissions to determine the fraction of X-ray luminosity. For a standard disk, the 2500{\AA} monochromatic luminosity ($L_{2500}$) can be estimated by assuming that the disk emission spectra $L_{\text{D,}\nu}\propto \nu^{-\gamma_{0}}$, where $\gamma_{0}=-1/3$. We can calculate $L_{2500}$ as

\begin{equation}
L_{2500}=\frac{L_{_{\rm D}}(1-\gamma_{0})}{\nu_{_{2500}}}(\frac{\nu_{_{2500}}}{\nu_{\text{p}}})^{1-\gamma_{0}},
\label{UV luminosity}
\end{equation}
where $\nu_{_{2500}}$ is the frequency corresponding to 2500{\AA}, $\nu_{\text{p}}$ is the peak frequency of the big blue bump, and the total disk luminosity is $L_{_{\rm D}}=\varepsilon L_{\rm Edd}=GM_{\rm BH}\dot{M}/6r_{\text{s}}$. The $\nu_{\text{p}}$ is determined by the maximum temperature of the thin disk and is estimated as (Lusso \& Risaliti 2017)
\begin{equation}
\nu_{\text{p}}\simeq7.2\times10^{17}(\frac{M_{\text{bh}}}{M_{\odot}})^{-1/4}(\frac{\dot{M}}{\dot{M}_{\text{Edd}}})^{1/4} \text{ Hz}.
\label{peak frequency}
\end{equation}
The nonlinear correlation between the 2 keV ($L_{2\text{keV}}$) and 2500{\AA} monochromatic luminosity is proposed as $\text{log}(L_{2\text{keV}})=0.642\text{log}(L_{2500})+6.965$ (Lusso \& Risaliti 2016), which is used to determine the 2 keV monochromatic luminosity. The corona emission spectra are assumed to be $L_{\text{cor},\nu}\propto \nu^{-\alpha_{\text{X}}}$, where $\alpha_{\text{X}}=0.9$ (Young et al. 2010). Then, the corona luminosity is calculated by
\begin{equation}
L_{_{\text{cor}}}=\frac{L_{2\text{keV}}(\nu^{1-\alpha_{\text{X}}}_{2}-\nu^{1-\alpha_{\text{X}}}_{1})}{(1-\alpha_{\text{x}})}\nu_{\text{X}}^{\alpha_{\text{X}}},
\label{UV luminosity}
\end{equation}
where $\nu_1$ and $\nu_2$ correspond to the frequency of 0.1 and 40 keV photon energy. Therefore, we obtain the fraction of X-ray luminosity $f_{\text{X}}=L_{\text{cor}}/L_{\text{D}}$. Here, the radiation temperature ($T_{\rm X}$) of the corona is set to be $10^8$ K. It is also noted that our simulations include only the contribution of X-ray luminosity in ionizing material rather than the effect of X-ray spectral shape. The X-ray spectral shape may play a role in radiation-driving fast outflows (Chartas et al. 2009b). In the future, the role of X-ray spectral shape should be studied in driving fast outflows.

\begin{figure}
\includegraphics[width=.45\textwidth]{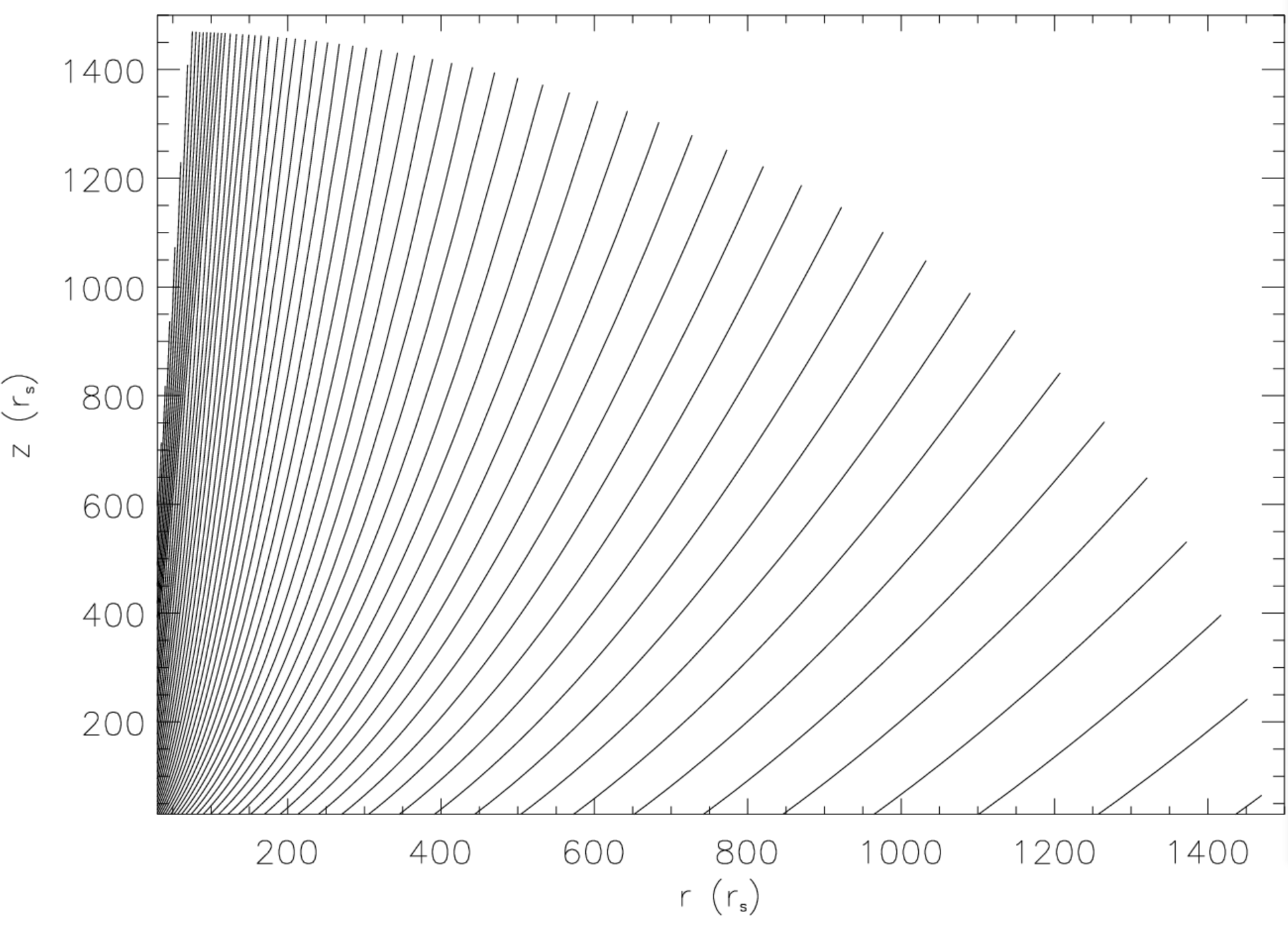}

\ \centering \caption{Initial magnetic field lines on the r-z plane when $\alpha_0$=5.0. }

 \label{fig1}
\end{figure}

\begin{figure}
\includegraphics[width=.45\textwidth]{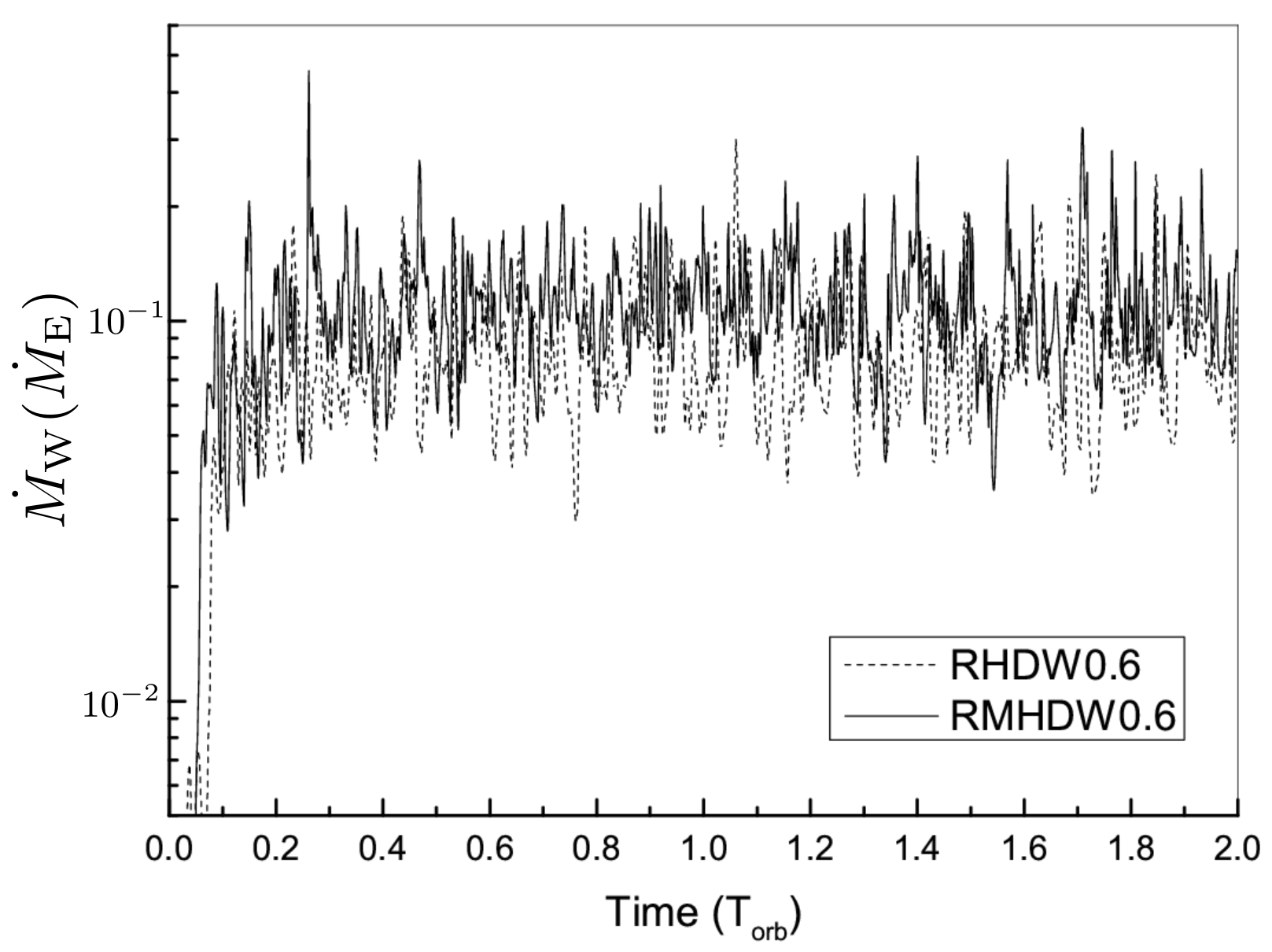}

\ \centering \caption{Time evolution of mass outflow rate at the outer boundary for runs RMHDW0.6 and RHDW0.6. In this figure, the units of mass outflow rate are Eddington accretion rate ($\dot{M}_{\rm E}=10 L_{\rm Edd}/c^2$).}

 \label{fig2}
\end{figure}

\subsection{Magnetic Field Configuration} \label{sec:MF}
Initially, a vector potential ${\bf A}\equiv(0,0,{\bf A}_{\phi})^{\rm T}$ is introduced to determine the magnetic field. In the spherical coordinate, the toroidal component ($
{\bf A}_{\phi}$) is written as
\begin{equation}
\begin{aligned}
{\bf A}_\phi=\frac{\Phi_{0}}{rsin(\theta)}\frac{r_{\text{in}}^2}{\alpha_0^2}[\sqrt{(\alpha_0\frac{r}{r_{\text{in}}}\sin(\theta))^2+(1+\alpha_0\frac{r}{r_{\text{in}}}\cos(\theta))^2}\\
-|1+\alpha_0\frac{r}{r_{\text{in}}}\cos(\theta)|],
\label{vectorpotential}
\end{aligned}
\end{equation}
where $r_{\text{in}}$ is the inner boundary of computation region and $\Phi_{0}$ and $\alpha_0$ are two parameters to determine the strength and inclination of initial magnetic field, respectively. The poloidal field components  (${\bf B}_{\rm p}\equiv{B}_r {\bf e}_r+{B}_\theta {\bf e}_\theta$) are given by $B_{r}=\frac{1}{r sin(\theta)}\frac{\partial({\bf A}_\phi sin(\theta))}{\partial \theta}$ and $B_{\theta}=-\frac{1}{r}\frac{\partial({\bf A}_\phi r)}{\partial r}$, which keeps $\nabla \cdot {\bf B}_{\rm p}=0$. The toroidal field component (${B}_\phi$) is initially zero. At the plane of $\theta=\pi/2$, We have
\begin{equation}
B_{r}(r,\theta=\pi/2)=\frac{\Phi_{0} r_{\text{in}}}{\alpha_0 r}[1-\frac{1}{\sqrt{1+(\alpha_0 \frac{r}{r_{\text{in}}})^2}}],
\end{equation}
\label{B_r}
and
\begin{equation}
B_{\theta}(r,\theta=\pi/2)=-\frac{\Phi_{0}}{\sqrt{1+(\alpha_0\frac{r}{r_{\text{in}}})^2}}.
\label{B_theta}
\end{equation}
The magnetic field  configuration is similar to that proposed by Blandford \& Payne (1982), who studied the self-similar disk wind.

$\beta_{0}=(8\pi P_{_{\text{s},}r_{\text{in}}})/B({r_{\text{in}},\pi/2})^2$ is used to scale the magnitude of the magnetic field at $r=r_{\text{in}}$. $P_{_{\rm s,}r_{\text{in}}}$ is the gas pressure at the disk surface at $r=r_{\text{in}}$. Then, we can obtain $\Phi_{0}^2=P_{_{\rm s,}r_{\text{in}}}[1+\alpha_0^2+\sqrt{1+\alpha_0^2}]/\beta_0$. Figure \ref{fig1} shows the initial magnetic field lines on the $r$--$z$ plane when $\alpha_0=5$.

\subsection{Model setup}
The computational domain does not include the thin disk and is located above the disk surface. The radial range covers 30$r_{\rm s}\leq r\leq$ 1500 $r_{\rm s}$. Initially, the gas in the computational domain is in the state of isothermal hydrostatic equilibrium in the vertical direction. The initial density distribution reads as
\begin{equation}
\rho(r,\theta)=\rho_{_{\rm d}} {\rm exp}(-\frac{GM_{_{\rm BH}}}{2c^2_{\rm s,\rm d}r(1-r_{\rm s}/r)^2{\rm tan}^2(\theta)}),
\label{density_distribution}
\end{equation}
where $\rho_{_{\rm d}}$ and $c_{\rm s, d}$ are the density and the isothermal sound speed at the disk surface, respectively. The isothermal sound speed $c_{\rm s, d}$ is determined by the effective temperature of disk surface, $T_{\rm eff}(r_{\rm D})=(\pi I_{\rm D}(r_{\rm D})/\sigma)^{\frac{1}{4}}$. The initial temperature $T(r,\theta)$ is set to be $T_{\rm eff}(r_{\rm D})$. For the initial velocities, ${\bf v}_r(r,\theta)$ and ${\bf v}_{\theta}(r,\theta)$ are set to be null. The rotational velocity is given as ${\bf v}_\phi(r,\theta)=\sqrt{GM_{_{\rm BH}}/r}{\rm sin}(\theta)r/(r-r_{\rm s})$, which meets the equilibrium between the BH gravity and the centrifugal force. We set the BH mass to be $10^8$ $M_{\odot}$.

We employ a nonuniform grid to discretize the computational domain. Except for run RMHDW0.6h, the computational region is divided into 144$\times$160 zones. The 144 zones with the radial size ratio, $(\bigtriangleup r)_{i+1} / (\bigtriangleup r)_{i} = 1.04$, are distributed in the $r$ direction and the smallest radial size is $\Delta r=0.20802$$r_{\rm s}$ at the inner boundary. The 16 zones are uniformly distributed over the angular range of $0^{\rm o}$--$15^{\rm o}$ while the 144 zones are nonuniformly distributed over the angular range of $15^{\rm o}$--$90^{\rm o}$. The angular size ratio is $(\bigtriangleup \theta)_{j+1} / (\bigtriangleup \theta)_{j} = 0.970072$ and then the smallest angular size is $\Delta \theta =0^{\rm o}.02358939$ at $\theta=90^{\rm o}$. For testing the effect of resolution, run RMHDW0.6h has higher resolution. For run RMHDW0.6h, the smallest radial size is $\Delta r=0.02306$$r_{\rm s}$ at the inner boundary, while the smallest angular size is $\Delta \theta=0^{\rm o}.00854799$ at $\theta=90^{\rm o}$. The outflow boundary condition is applied at the inner and outer radial boundaries and the axially symmetric boundary condition is employed at the pole (i.e. $\theta=0$). At the equator (i.e. $\theta=90^{\rm o}$), density and temperature are set to be $\rho_{\rm d}$ and $T_{\rm eff}(r_{\rm D})$ at all times, the radial and vertical velocities are set to be null, and the rotating velocity is set to the Kepler velocity.

Equations (1)--(4) are numerically solved by the following steps. We firstly use PLUTO code to numerically solve the HD/MHD equations by an HLLC Riemann solver (Mignone et al. 2007; 2012). As an extended version of the HLL solver, the HLLC Riemann solver can achieve high numerical resolution (Li 2005). Then, the radiation force in equation (3) is treated as an external force term to explicitly update velocities. Finally, the net cooling rate in equation (4) is used to implicitly update the gas temperature (i.e. internal energy of gas) using the bisection method.

\begin{figure*}

\includegraphics[width=.49\textwidth]{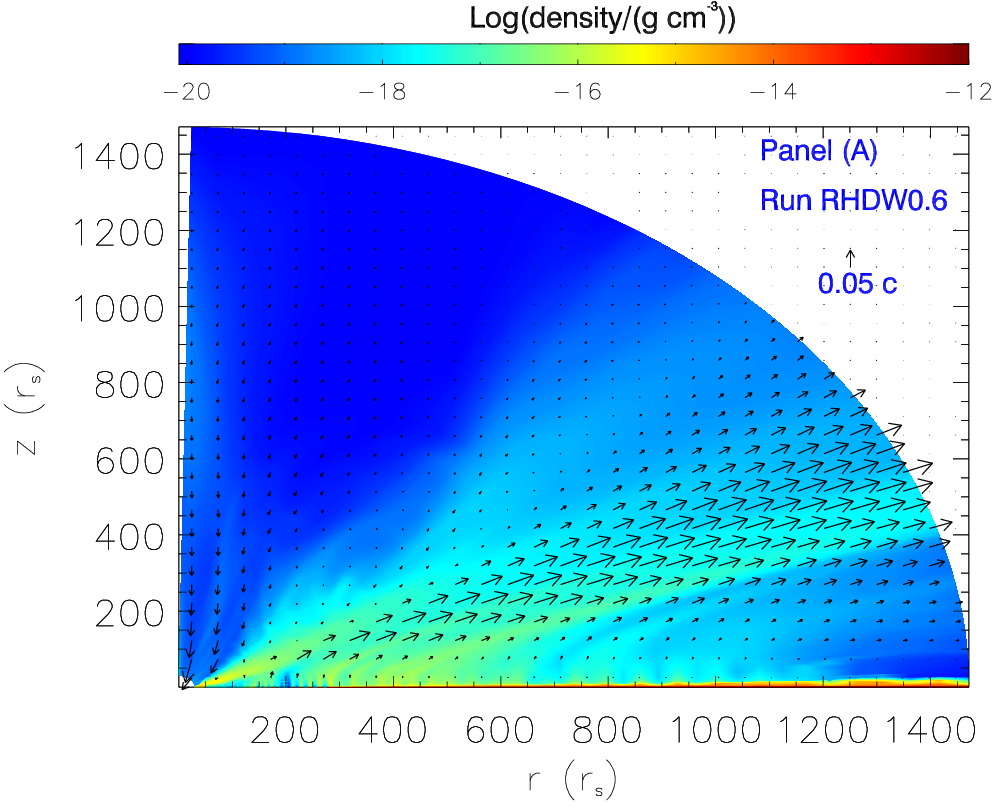}
\includegraphics[width=.49\textwidth]{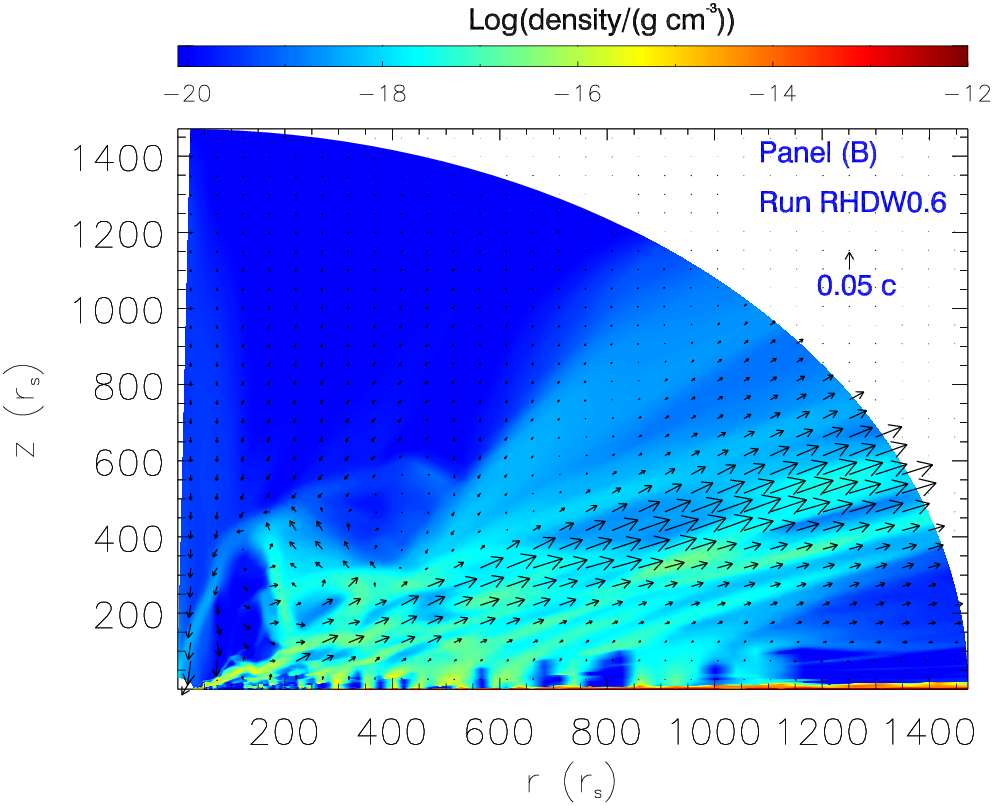}

\includegraphics[width=.49\textwidth]{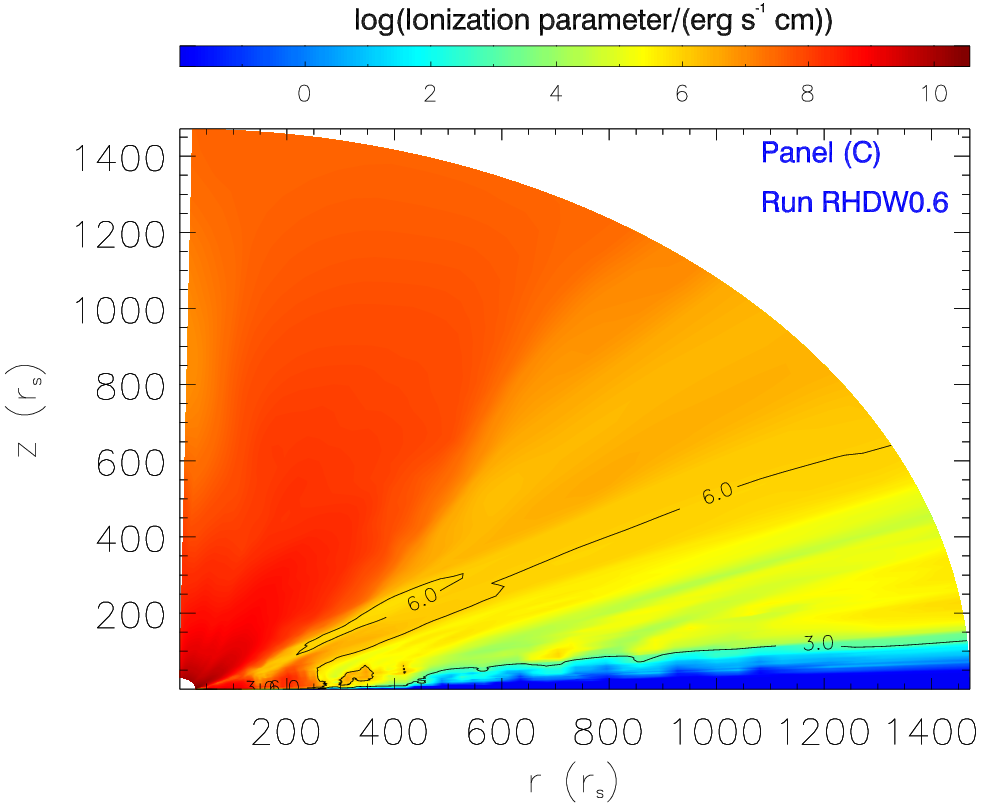}
\includegraphics[width=.49\textwidth]{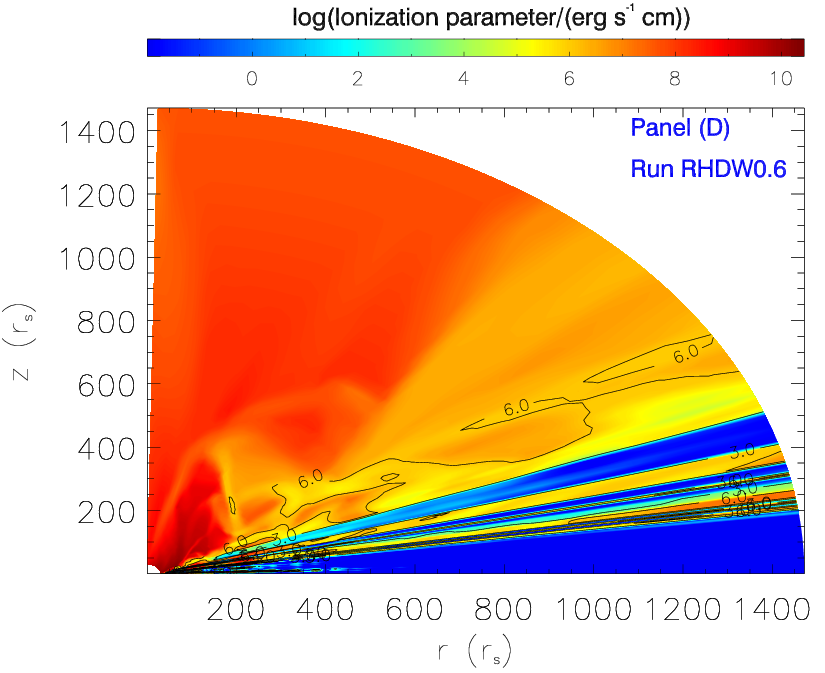}

\includegraphics[width=.49\textwidth]{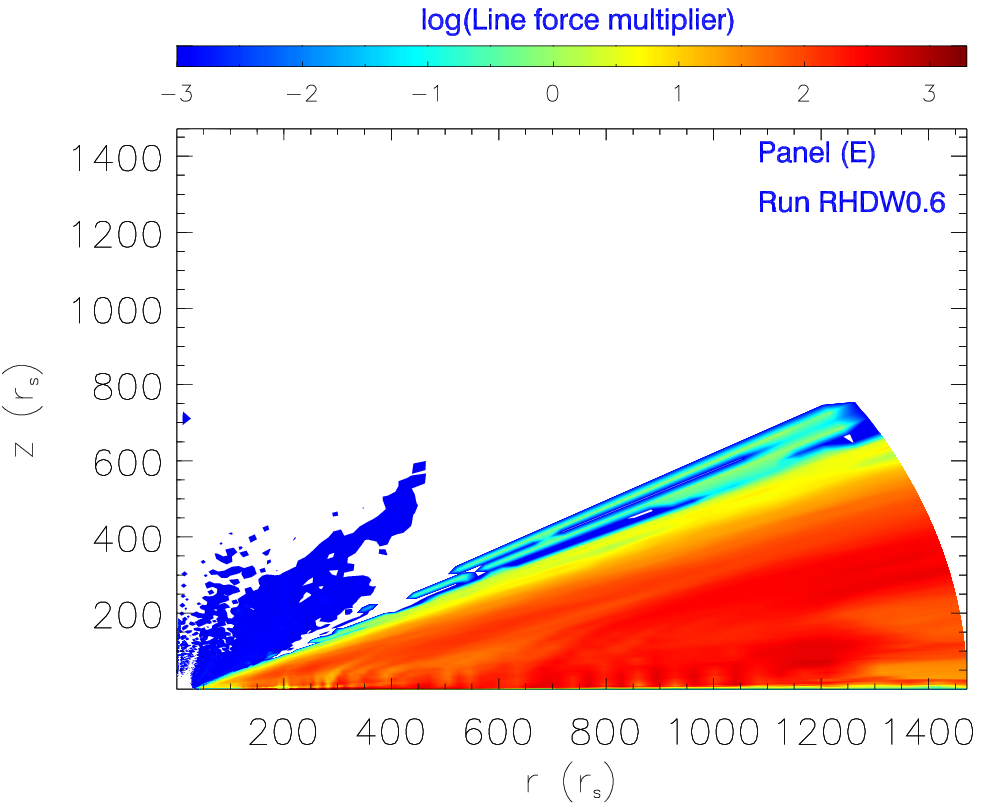}
\includegraphics[width=.49\textwidth]{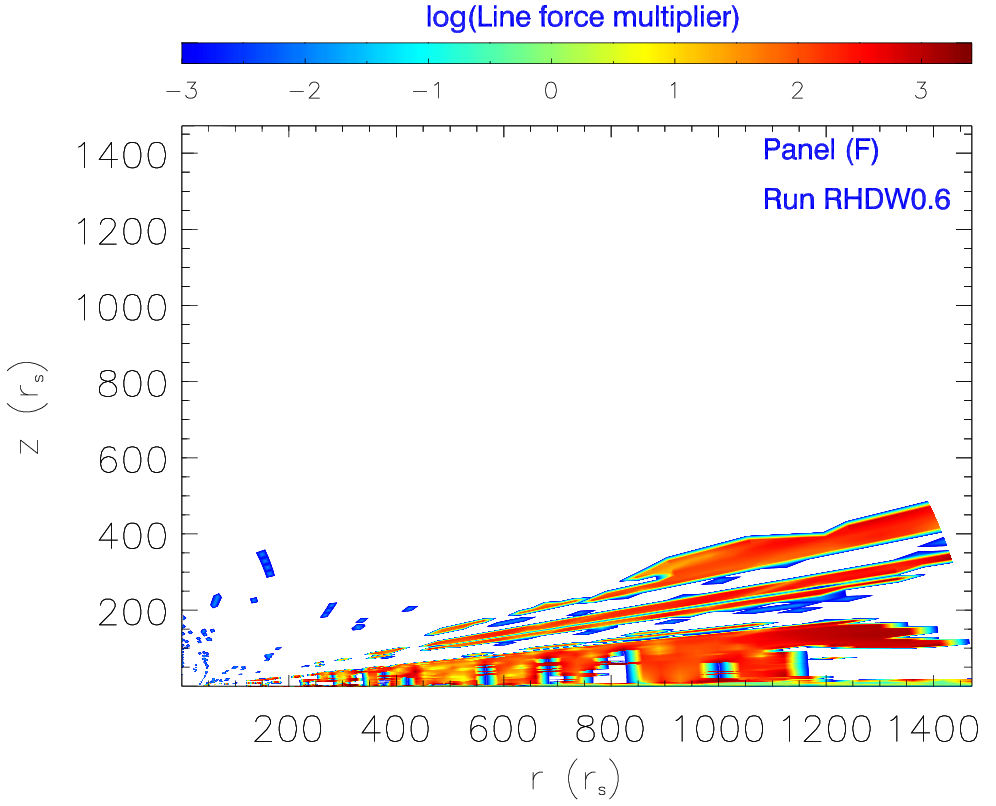}

\ \centering \caption{Two-dimensional distribution of a series of variables for run RHDW0.6. Left columns give the time-averaged results on the range of 0.5--1.0 $T_{\rm orb}$ and right columns give a snapshot at $t=1.0$ $T_{\rm orb}$. Panels (A) and (B) show the density distribution and poloidal velocity; panels (C) and (D) show the ionization parameter $\xi$; panels (E) and (F) show the line-force multiplier $\mathcal{M}$.}

\label{fig3}
\end{figure*}

\begin{figure*}
\includegraphics[width=.49\textwidth]{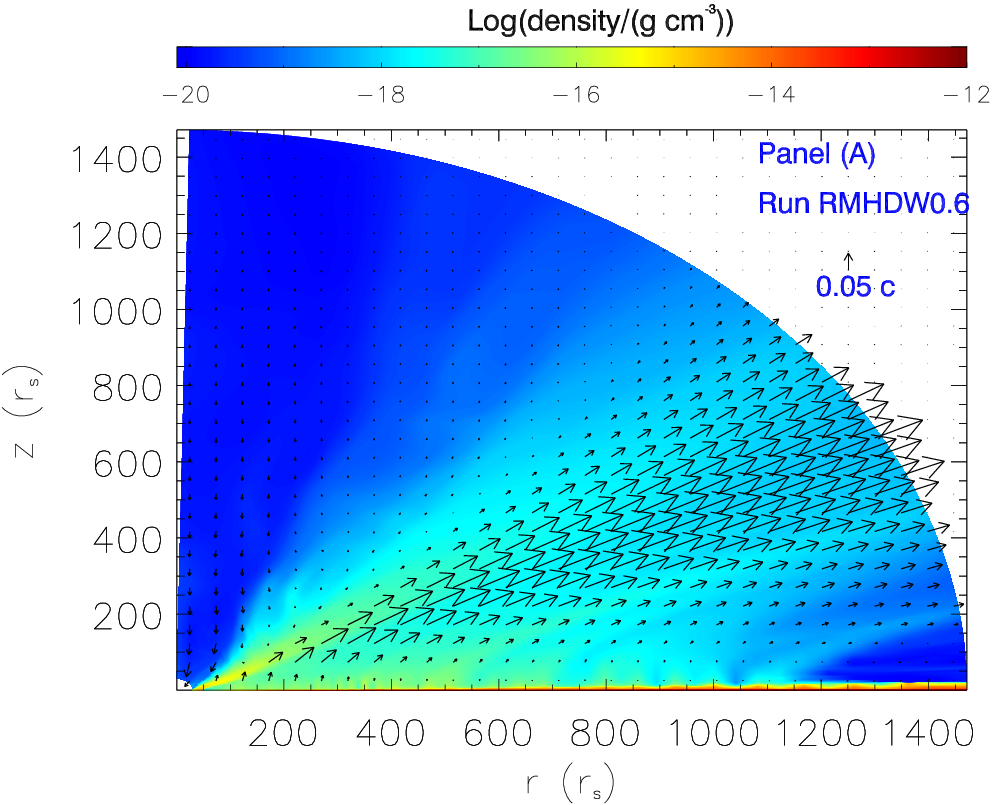}
\includegraphics[width=.49\textwidth]{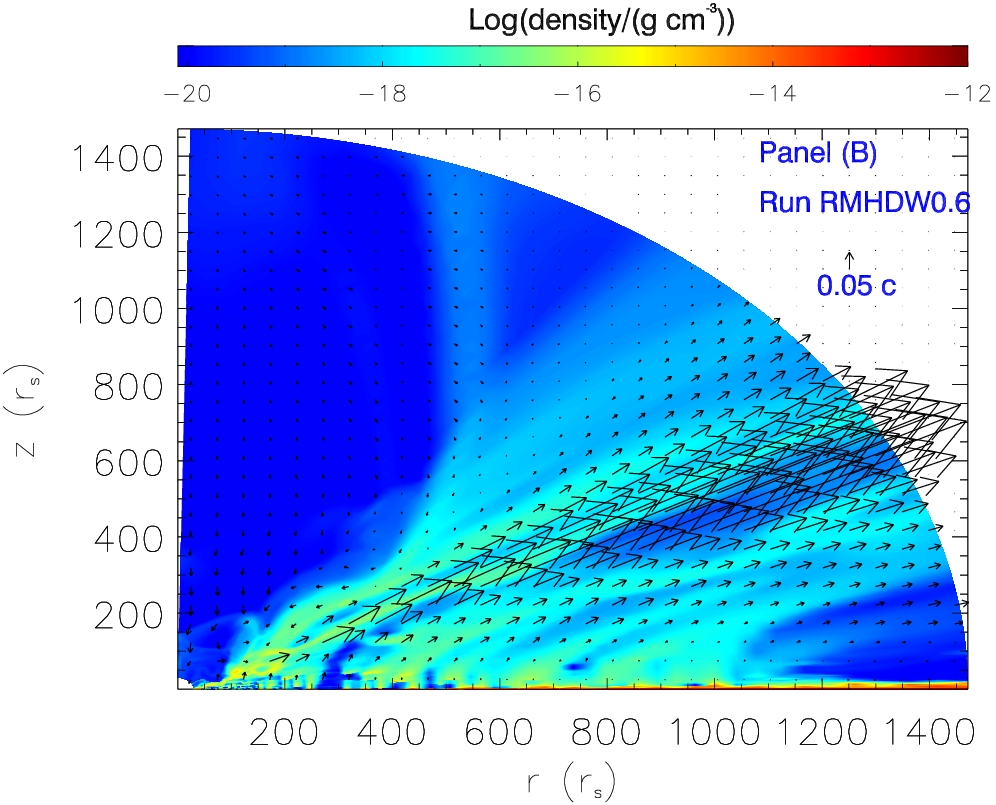}

\includegraphics[width=.49\textwidth]{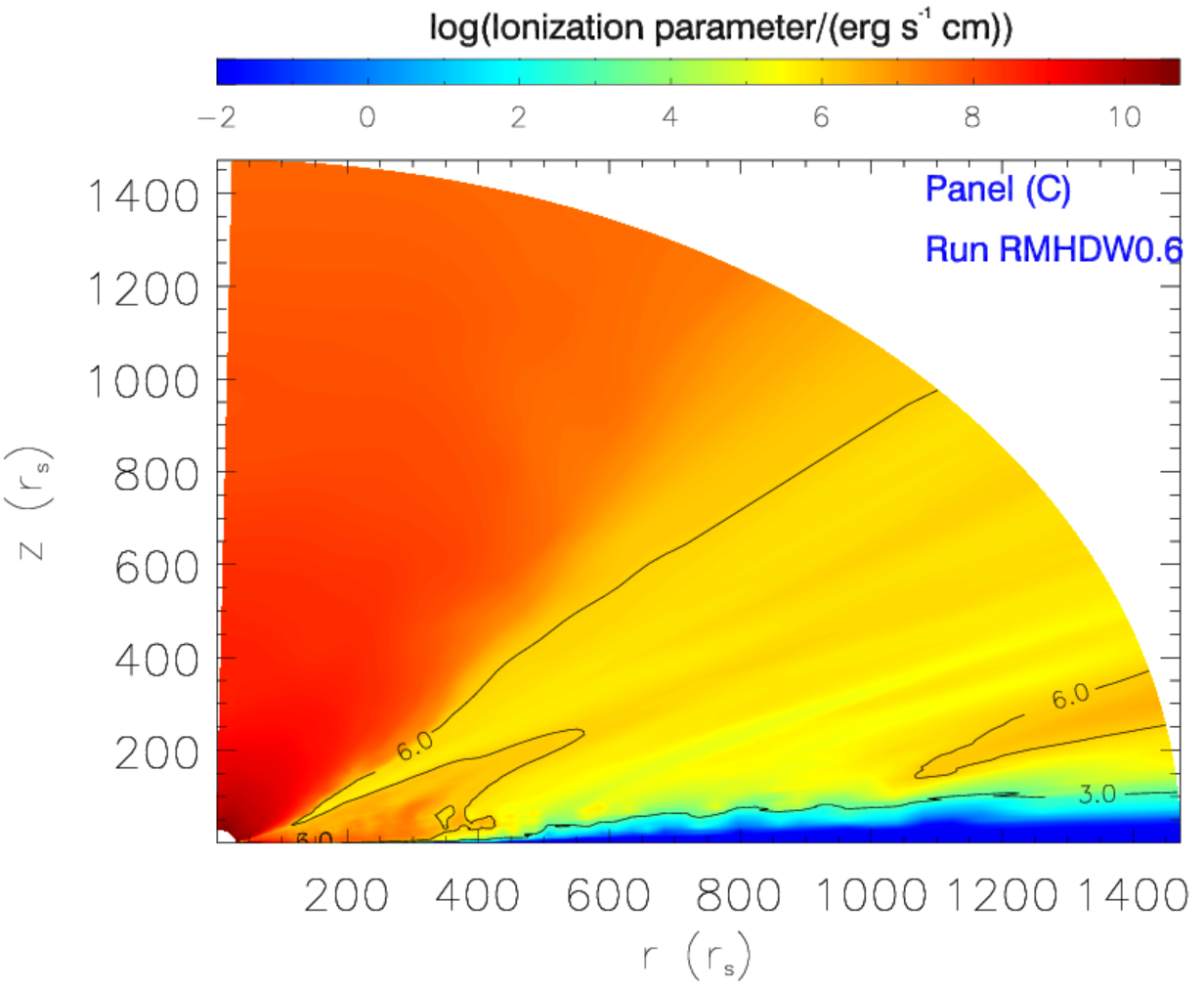}
\includegraphics[width=.49\textwidth]{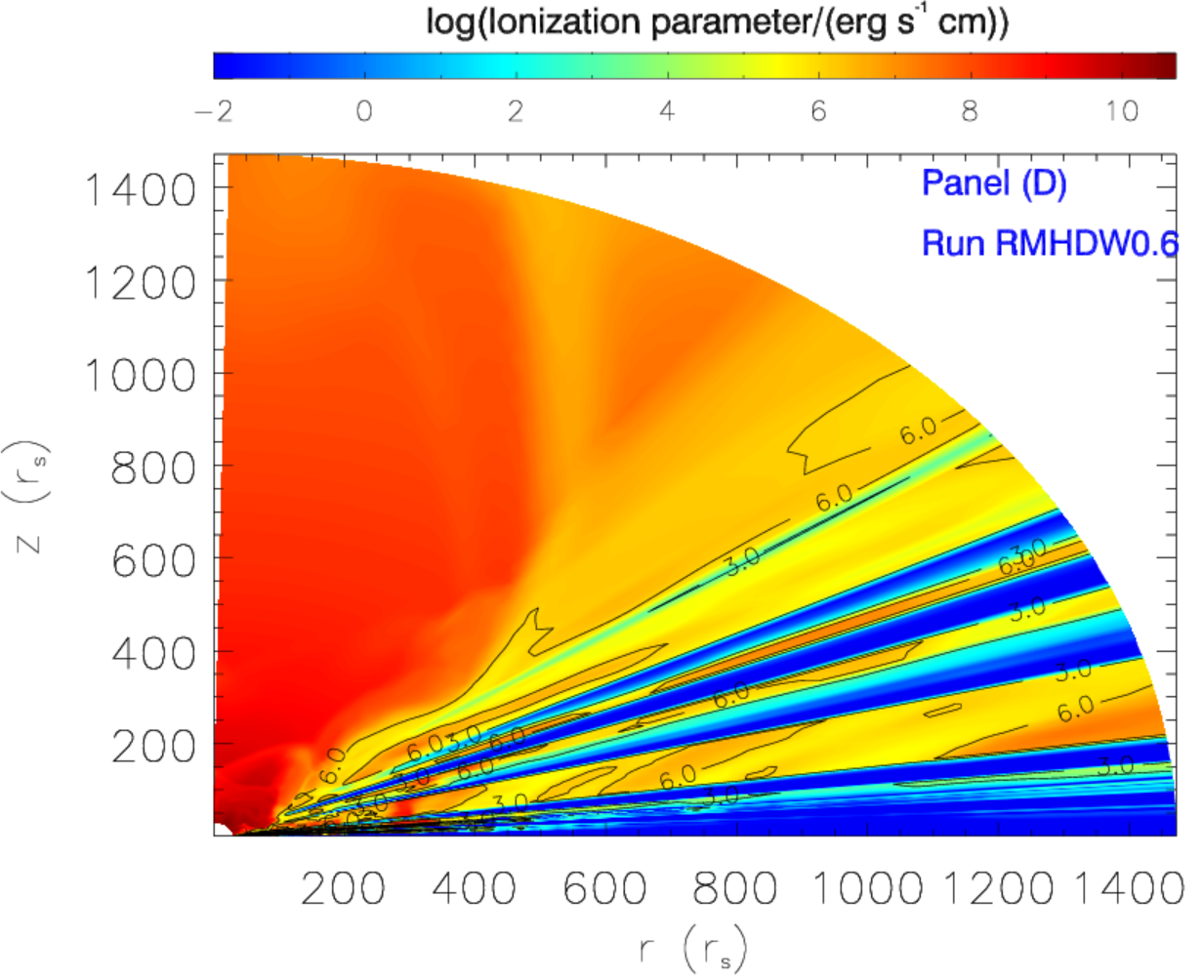}

\includegraphics[width=.49\textwidth]{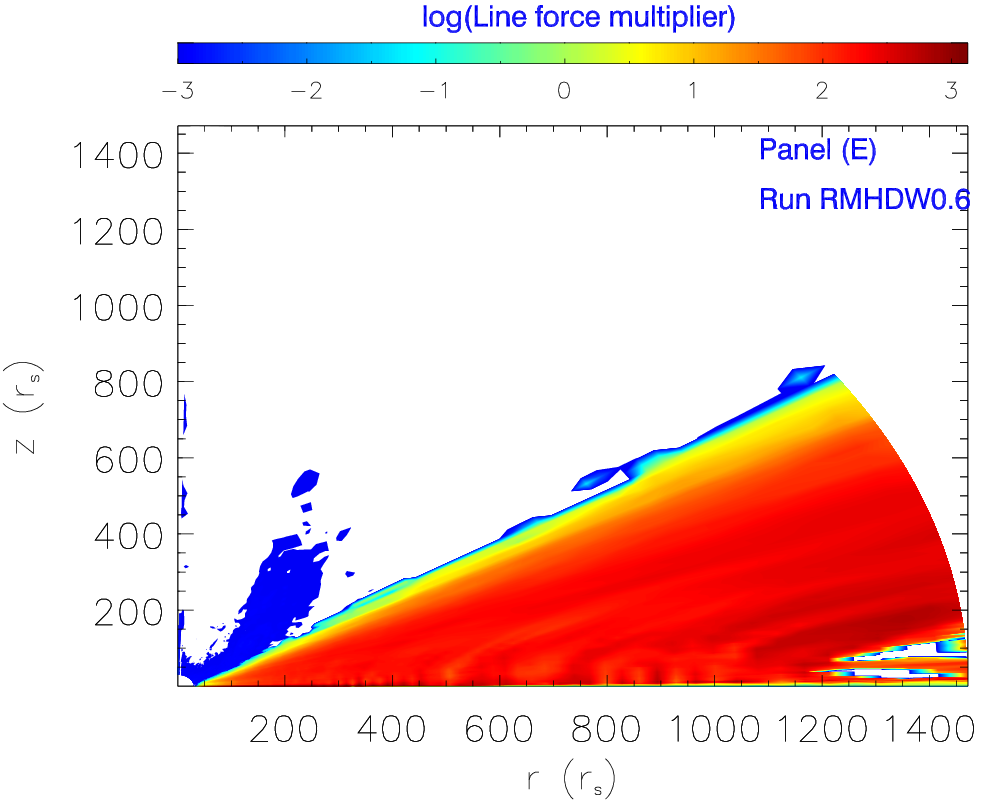}
\includegraphics[width=.49\textwidth]{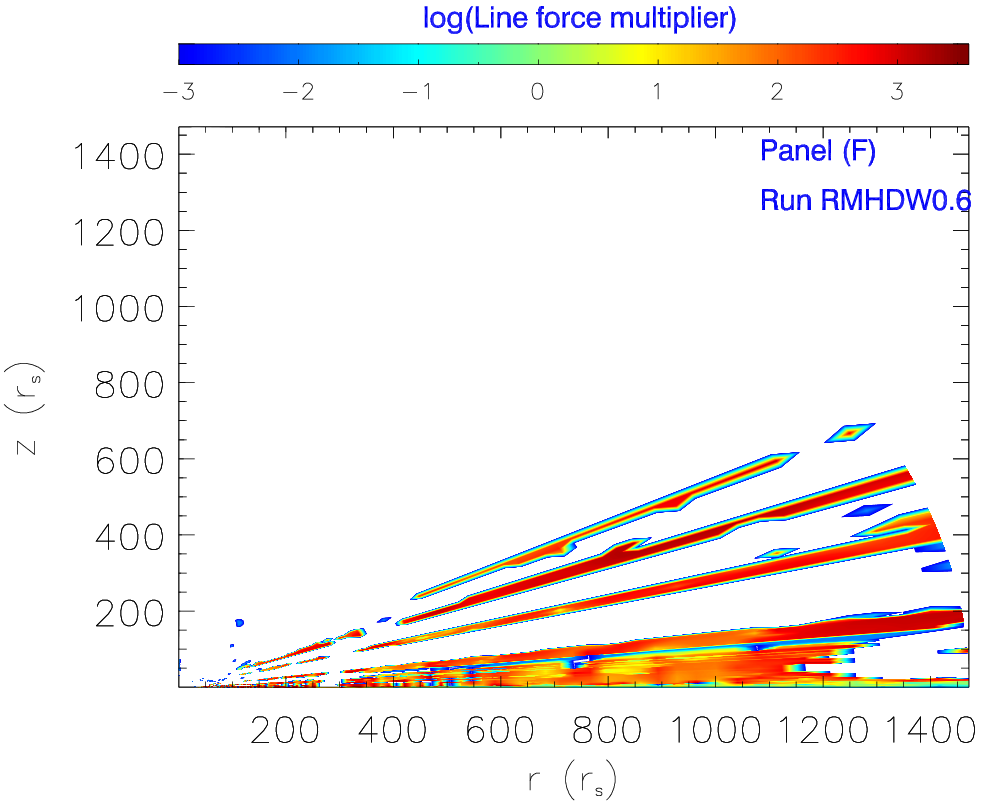}

\ \centering \caption{Two-dimensional distribution of a series of variables for run RMHDW0.6. Panels (A)--(F) are the same as the panels of Figure \ref{fig3}.}

\label{fig4}
\end{figure*}

\begin{figure*}
\includegraphics[width=.32\textwidth]{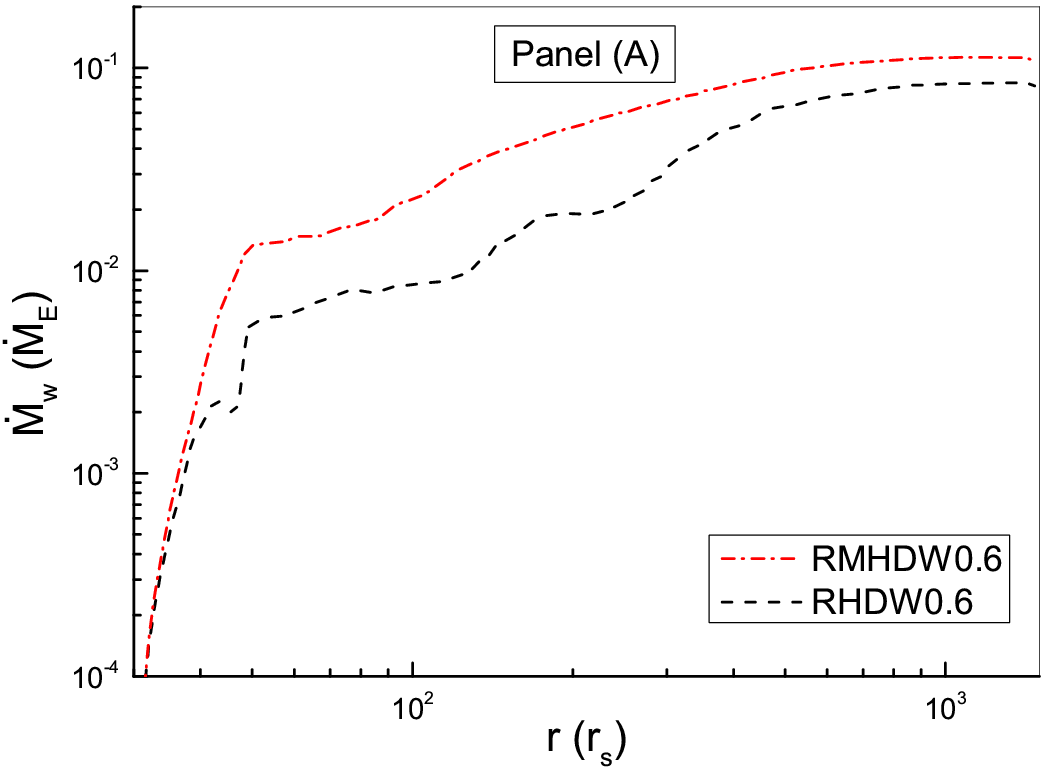}
\includegraphics[width=.32\textwidth]{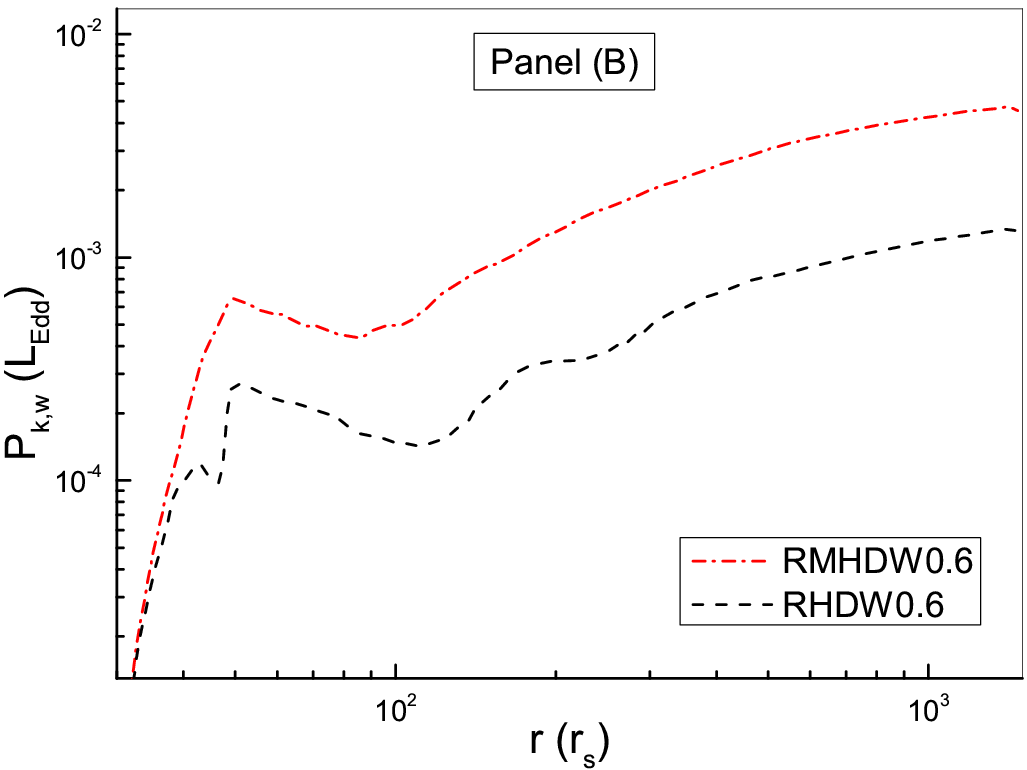}
\includegraphics[width=.32\textwidth]{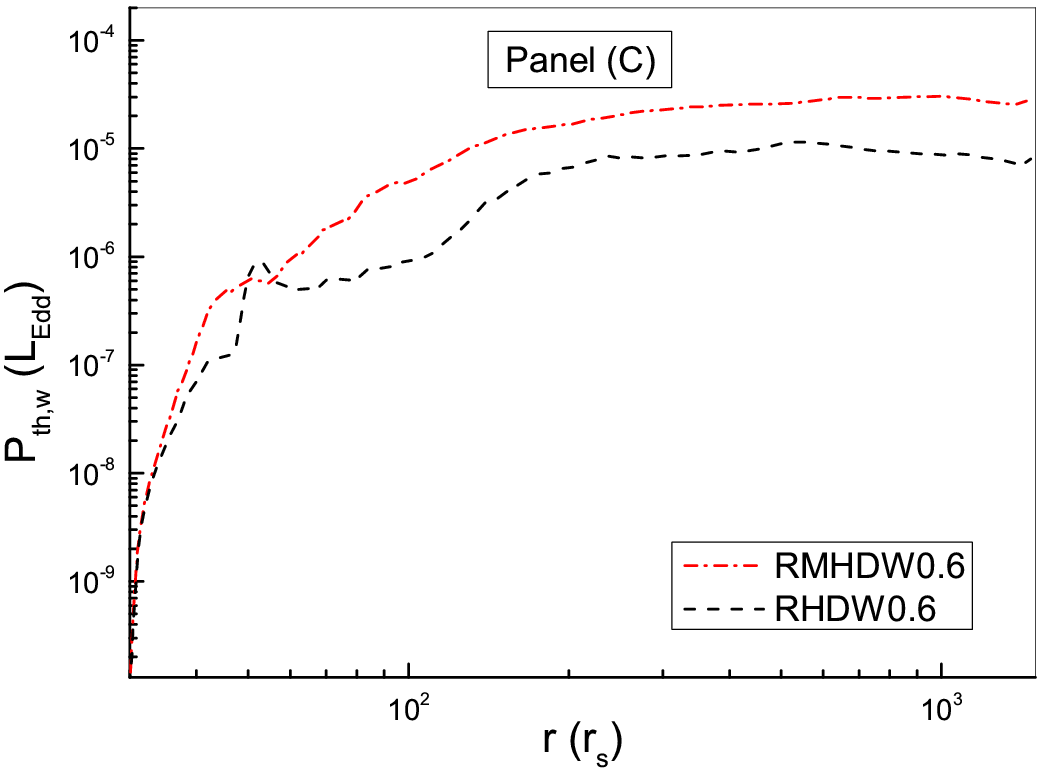}

\ \centering \caption{Radial dependence of time-averaged quantities. Panel (A): mass outflow rate in units of Eddington accretion rate ($\dot{M}_{\rm E}=10 L_{\rm Edd}/c^2$); Panel (B): kinetic power of winds; panel (C): thermal energy flux carried by outflows. }
\label{fig5}
\end{figure*}

\begin{figure*}
\includegraphics[width=.32\textwidth]{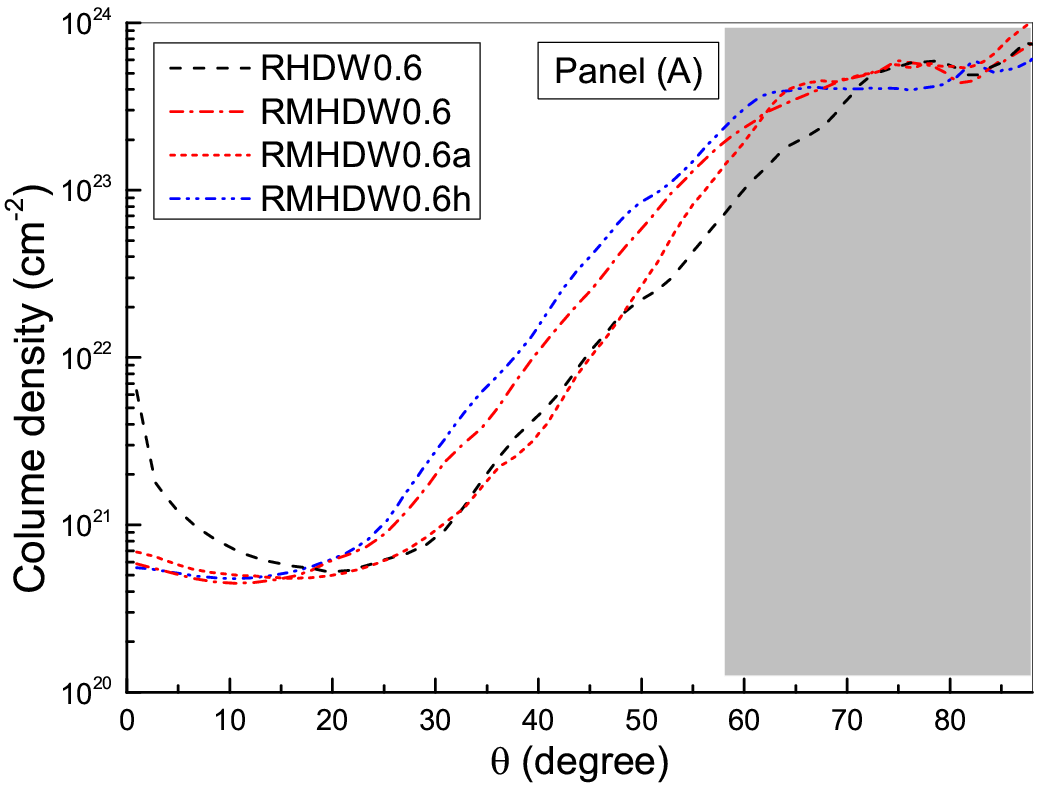}
\includegraphics[width=.32\textwidth]{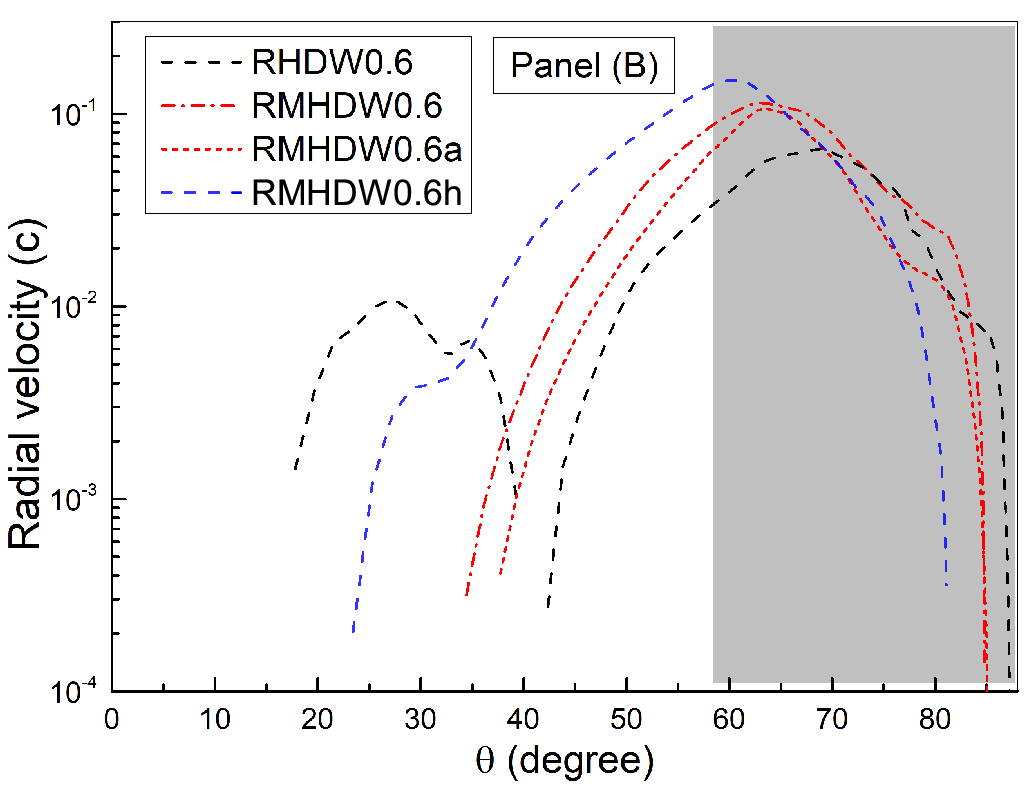}
\includegraphics[width=.32\textwidth]{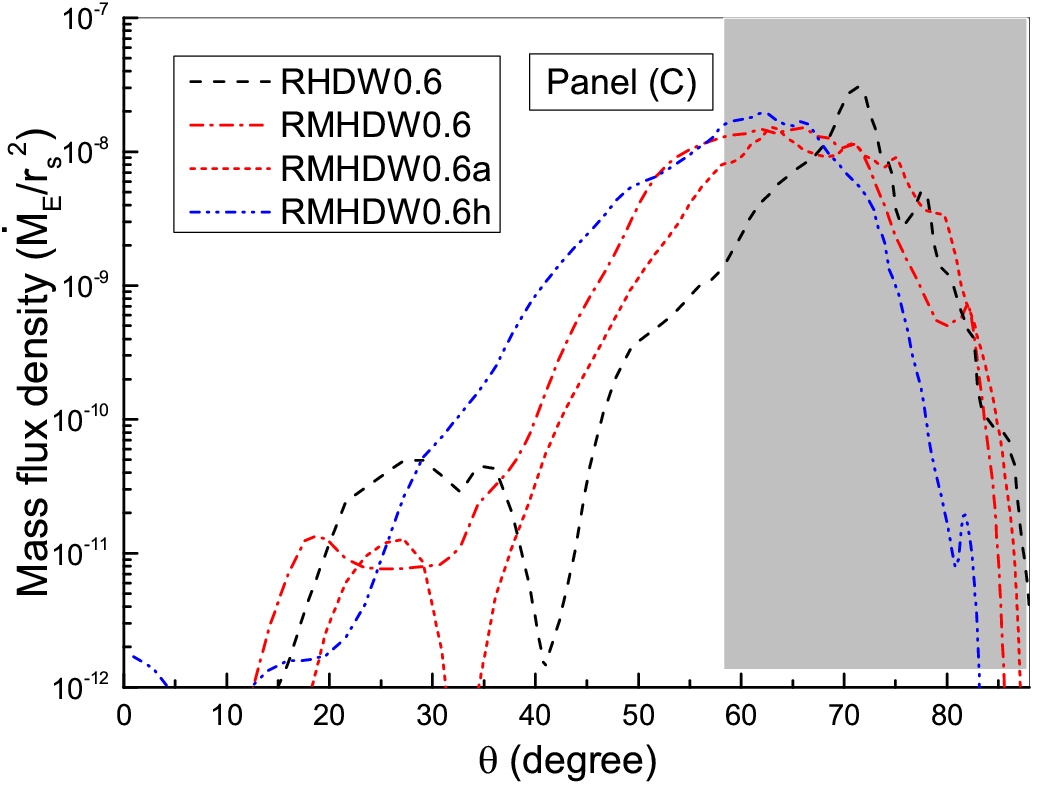}
%\scalebox{0.36}[0.4]{\rotatebox{0}{\includegraphics[bb=40 20 500 360]{fig6_a.eps}}}
%\scalebox{0.36}[0.4]{\rotatebox{0}{\includegraphics[bb=40 20 500 360]{fig6_b.eps}}}
%\scalebox{0.36}[0.4]{\rotatebox{0}{\includegraphics[bb=40 20 500 360]{fig6_c.eps}}}

\ \centering \caption{Angular profiles of a variety of time-averaged variables. Panel (A): column density; Panel (B): radial velocity at the outer boundary; Panel (C): mass flux density at the outer boundary. Gray belt denotes the angular range where the force multiplier of the line force is not negligible. }
\label{fig6}
\end{figure*}

\section{Results} \label{sec:Results}
Our runs are summarized in Table 1, where hydrodynamic simulation is implemented when $\beta_{0}$ is infinity. Table 1 gives the time-averaged properties of winds at the outer boundary, such as the mass outflow rate ($\dot{M}_{\rm w}$), the momentum flux ($P_{\rm w}$) as well as the kinetic ($P_{\rm k,w}$) and thermal energy ($P_{\rm th,w}$) fluxes. These values are obtained by time averaging the time range of 0.5--1.0 $T_{\rm orb}$, where $T_{\rm orb}$ is the Keplerian orbital period at the outer boundary. A quasi-steady state can be quickly reached in all of the runs. Figure \ref{fig2} shows the time evolution of the mass outflow rate at the outer boundary for runs RHDW0.6 and RMHDW0.6. For run RHDW0.6, we set $\beta_{0}=\infty$ and then do not take into account the effect of magnetic field. For run RMHDW0.6, we $\beta_{0}=100$ so that the magnetic field is weak. For testing the effect of X-ray strength (i.e. $f_{\rm X}$) and grid resolution on the wind properties, we further run RMHDW0.6a and RMHDW0.6h. In RMHDW0.6a, $f_{\rm X}$ is set to be two times than that that in MHDW0.6. Table 1 shows that the magnetic field is helpful in increasing the mass outflow rate and power of winds, though the initial magnetic field is weak.

\subsection{Properties of Winds} \label{subsec:Winds}

Runs RHDW0.6 and RMHDW0.6 are used as two examples to show the two-dimensional structure of solutions. Their density, poloidal velocity, ionization parameter, and line-force multiplier are shown in Figures \ref{fig3} and \ref{fig4}. In the two figures, the left columns give the time-averaged results of the time interval of 0.5--1.0 $T_{\rm orb}$ while the right columns give snapshots at $t=1.0$ $T_{\rm orb}$. The time interval of averaging is much longer than the dynamic time scale of the gas from the inner boundary to the outer boundary. This ensures that the time interval of averaging does not occur for a time period for which the averaged results happened to be higher or lower than averaged. We also sample data at snapshots, and the data are around the averaged values. As shown in panels (A) and (B) in Figure \ref{fig3}, winds are generated from the disk surface within the 600 $r_{s}$ and then blown away from the disk surface. Comparing Figures \ref{fig3} and \ref{fig4}, we find that the magnetic field does not significantly change the two-dimensional structure of winds. In run RMHDW0.6, the winds have higher velocity than that in run RHDW0.6.

The force multiplier $\mathcal{M}$ depends on the ionization parameter. When $\xi\gtrsim 10^2$ erg s$^{-1}$ cm, the maximum value of $\mathcal{M}$ is close to 1, and so the line force becomes negligible. When $\xi$ increases from 0 to $\sim3$ erg s$^{-1}$ cm, the maximum value of $\mathcal{M}$ increases gradually from $\sim2000$--$5000$ (Proga et al. 2000). In that case, the line force becomes very significant. From panels (A) and (E) of Figs. 3 and 4, we can see that almost all of the high-velocity winds undergo an angular range of line-force acceleration and they are efficiently accelerated within the angular range. Tombesi et al. (2011) pointed out that the ionization parameter of UFOs is very high and in the range of log($\xi$/(erg s$^{-1}$ cm))$\sim$3--6. UFOs cannot be directly driven by line force. Panels (D) and (F) show that the line force operates over the region like filaments on a snapshot. When the X-ray photons are shielded, the degree of ionization of the shielded gases becomes low and then the line force becomes effective. When a great number of X-ray photons can effectively pass through the gases on the inner region, the gas on the outer region is exposed to X-rays and then the line force becomes invalid due to the gas becoming highly ionized by the X-ray photons. The change in the gas distribution on the inner region causes the filaments' location to changes with time. The line-force driven winds exposed to X-rays form  high-ionization winds with high velocity. Therefore, the high-velocity winds are in a multiphase, i.e., low-ionization winds (log($\xi$/(erg s$^{-1}$ cm))$\sim$0--3) and high-ionization winds (log($\xi$/(erg s$^{-1}$ cm))$\sim$3--6), and then high-ionization winds could be UFOs.

\begin{figure*}
\includegraphics[width=.49\textwidth]{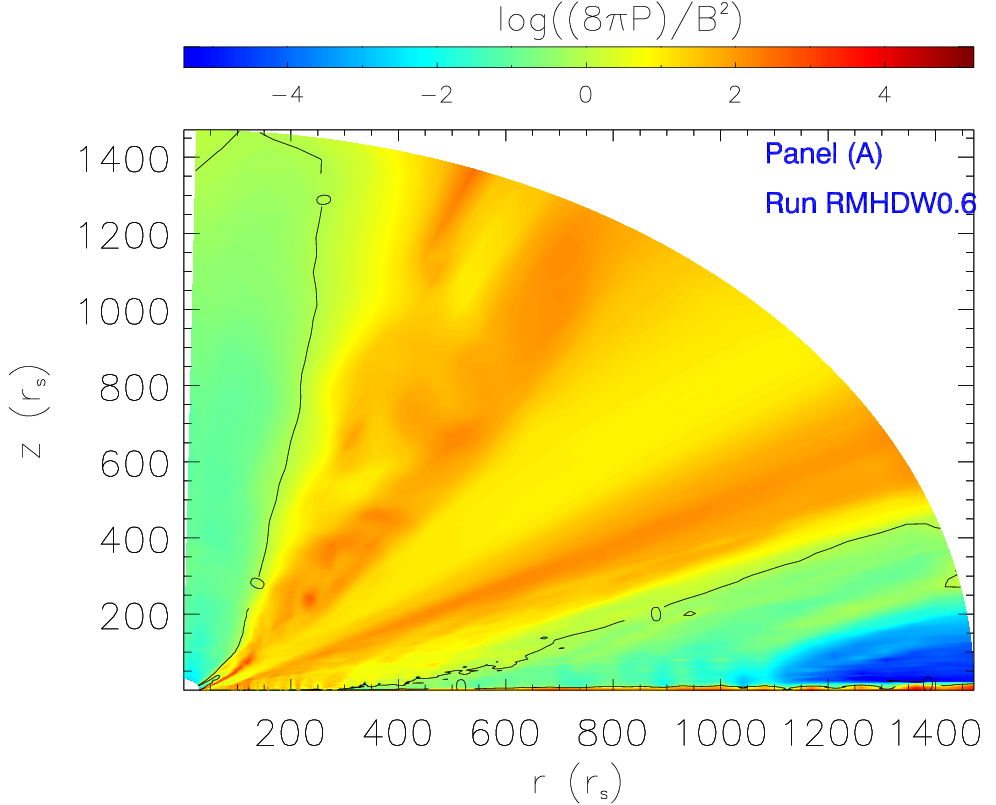}
\includegraphics[width=.49\textwidth]{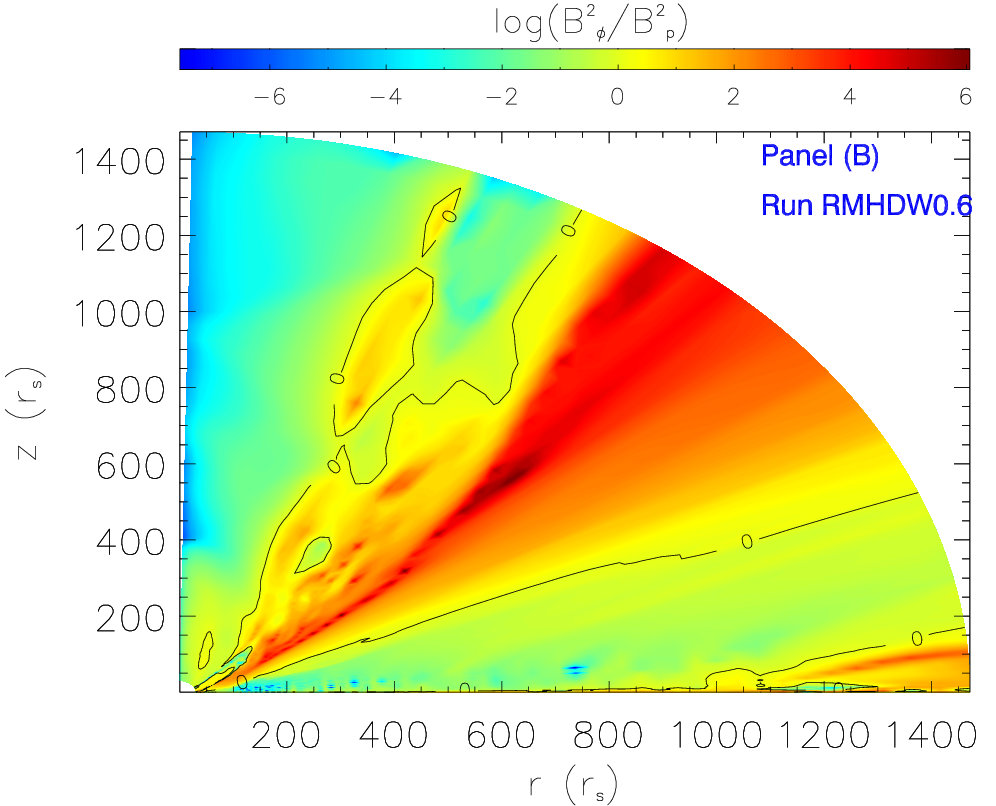}
\includegraphics[width=.49\textwidth]{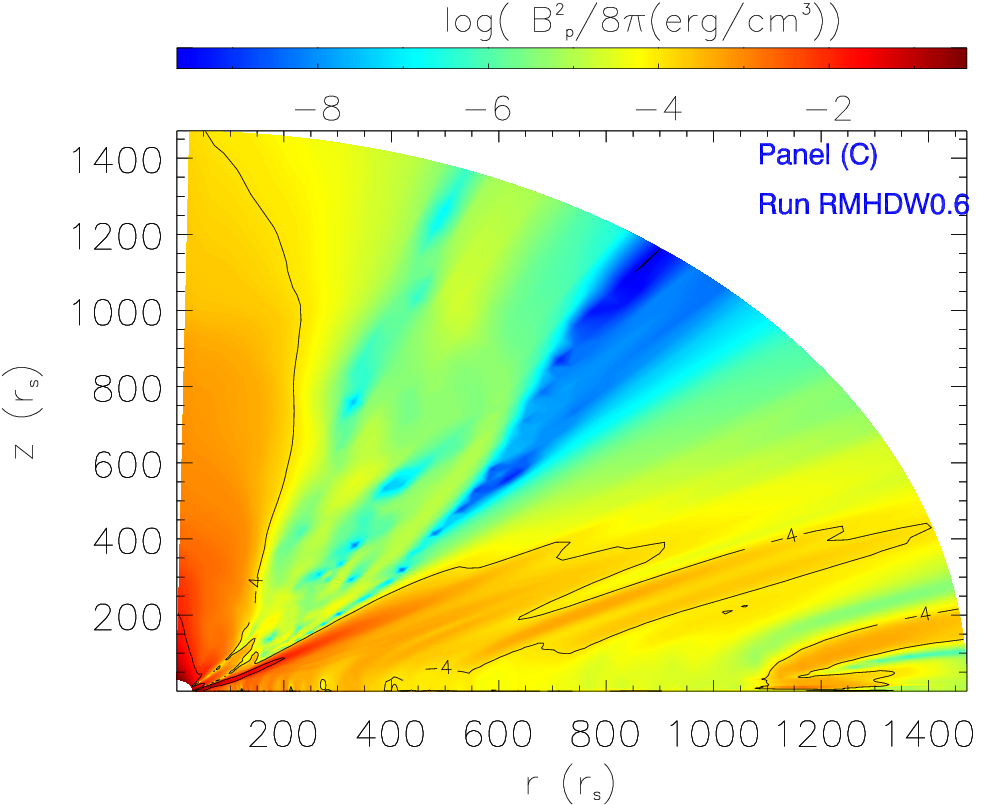}
\includegraphics[width=.49\textwidth]{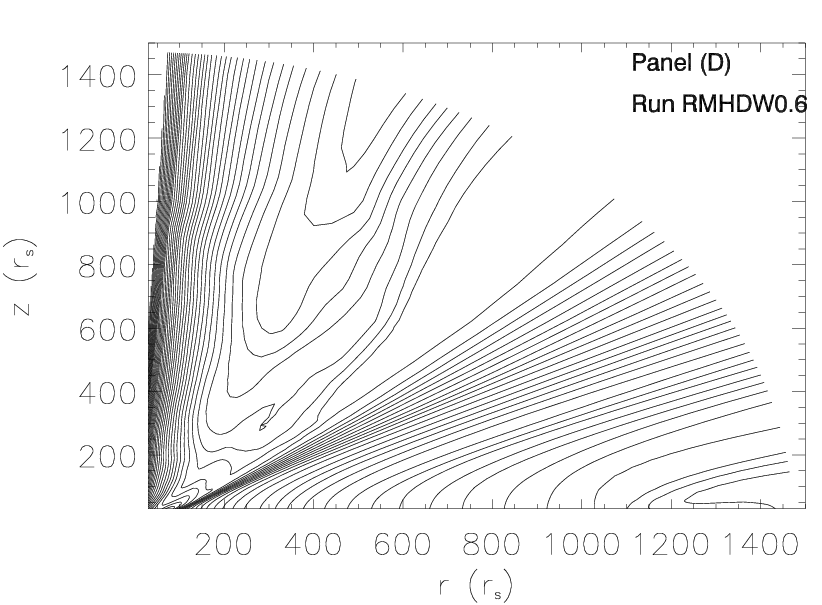}
%\scalebox{0.5}[0.55]{\rotatebox{0}{\includegraphics[bb=70 360 490 720]{fig7_RMHDW0.6_beta.ps}}}
%\scalebox{0.5}[0.55]{\rotatebox{0}{\includegraphics[bb=70 360 490 720]{fig7_RMHDW0.6_bphi2bp2.ps}}}
%\scalebox{0.5}[0.55]{\rotatebox{0}{\includegraphics[bb=70 360 490 720]{fig7_RMHDW0.6_bp2.ps}}}
%\scalebox{0.5}[0.55]{\rotatebox{0}{\includegraphics[bb=70 360 490 720]{fig7_RMHDW0.6_Maglines.ps}}}
\ \centering \caption{Structure of the magnetic field. Panel (A) shows the logarithm of the ratio of gas pressure ($P$) to magnetic pressure ($B^2/8\pi$). Panel (B) shows the logarithm of the ratio ($B^2_{\phi}/B^2_{\rm p}$) of toroidal ($B^2_{\phi}/8\pi$) to poloidal ($B^2_{\rm p}/8\pi$) magnetic pressure. Panel (C) shows the energy density of the poloidal magnetic field. Panel (D) shows the magnetic field lines on the $r$--$z$ plane. }

\label{fig7}
\end{figure*}

In the sense of time average, for run RHDW0.6, the line force becomes significant on the $\theta>64^{o}$ angular range and the 3$\lesssim$log($\xi$/(erg s$^{-1}$ cm))$\lesssim$6 region is distributed on the 65$^{\rm o}$--85$^{\rm o}$ angle range; for run RMHDW0.6, the line force becomes significant on the $\theta>58^{o}$ angular range and the 3$\lesssim$log($\xi$/(erg s$^{-1}$ cm))$\lesssim$6 region is distributed on the 57$^{\rm o}$--85$^{\rm o}$ angle range. The magnitude of the line force is weaker in the inner region ($<100$ $r_{\rm s}$) than in the outer region ($>100$ $r_{\rm s}$), because the inner-region gas is in the state of higher ionization compared with the outer-region gas. Initially, the gas at the disk surface rotates at the Keplerian and is in the state of force balance in the vertical direction. When an additional force, such as radiation force, is exerted on the gas, the gas may be blown off from the disk surface. In the inner region ($<100$ $r_{\rm s}$), the line force and Compton-scattering force blow off gas from the disk surface. The blown-off gas is further accelerated by the line force in the outer region ($>100$ $r_{\rm s}$).

In the following, we quantitatively analyze the the properties of winds. The properties of winds vary with radius. We calculate the radial dependence of the mass outflow rate ($\dot {M}_{\rm w} (r)$) and the kinetic ($P_{\rm k,w} (r)$) and thermal energy ($P_{\rm th,w} (r)$) carried by the outflow. These physical quantities are used to quantitatively describe the radial dependence of wind properties. $\dot {M}_{\rm w} (r)$, $P_{\rm k,w} (r)$, and $P_{\rm th,w} (r)$ are, respectively, given by
\begin{equation}
\dot {M}_{\rm w} (r)=4\pi r^2 \int_{\rm 0^\circ}^{\rm 89^\circ}
\rho \max (v_r, 0) \sin\theta d\theta,
\end{equation}

\begin{equation}
P_{\rm k} (r)=2\pi r^2 \int_{\rm 0^\circ}^{\rm 89^\circ} \rho
\max(v_r^3,0) \sin\theta d\theta,
\end{equation}
and
\begin{equation}
P_{\rm th} (r)=4\pi r^2 \int_{\rm 0^\circ}^{\rm 89^\circ} e
\max(v_r,0) \sin\theta d\theta.
\end{equation}
We time averaged 500 output files over the time range of $t=0.5$--1.0 $T_{\rm orb}$ and then obtained the time-averaged quantities. Figure \ref{fig5} shows the radial dependence of the time-averaged $\dot {M}_{\rm w} (r)$, $P_{\rm k,w} (r)$, and $P_{\rm th,w} (r)$. In figure 5, $\dot{M}_{\rm E}$ is defined as the Eddington accretion rate ($\dot{M}_{\rm E}=10 L_{\rm Edd}/c^2$). $\dot {M}_{\rm w} (r)$, $P_{\rm k,w} (r)$, and $P_{\rm th,w} (r)$ rapidly increase within $\sim$50 $r_{\rm s}$ and continue to increase outside 50 $r_{\rm s}$. $\dot {M}_{\rm w} (r)$ almost remaining constant outside 800 $r_{\rm s}$, while $P_{\rm th,w} (r)$ remains constant outside 200 $r_{\rm s}$. In the case with a magnetic field, such as run RMHDW0.6, the effect of the magnetic field enhances $\dot {M}_{\rm w} (r)$ outside $\sim$50 $r_{\rm s}$ and $P_{\rm k,w} (r)$.

Figure \ref{fig6} shows the angular dependence of winds at the outer boundary. In Figure \ref{fig6}, we plot the angular distribution of the column density ($N_{H}=\int^{1500r_{\rm s}}_{30r_{\rm s}}\frac{\rho(r,\theta)}{\mu m_{p}} dr$), the radial velocity, and the mass flux density. The gray belt in figure \ref{fig6} denotes the angular range where the force multiplier of the line force is significant. Close to the disk surface, the radial radiation flux significantly decreases. The radial component of the line force becomes neglected around the disk surface. As shown in Figure \ref{fig6}, the column density is a function of $\theta$. The column density increases from $\sim6\times10^{19}$ cm$^{-2}$ at $\theta=20^{\rm o}$ to $\sim8\times10^{23}$ cm$^{-2}$ at $\theta=88^{\rm o}$. In the region of $20^{\rm o} <\theta<72^{\rm o}$, the column density of run RMHDW0.6 is higher than that in the HD case, while their column density is comparable in the region of $72^{\rm o} <\theta<88^{\rm o}$. The column density is higher than $10^{22}$ cm$^{-2}$ when $\theta\gtrsim40^{\rm o}$ for run RMHDW0.6 ($\theta\gtrsim45^{\rm o}$ for run RHDW0.6). The mass flux density of winds in run RMHDW0.6 is much larger than that in run RHDW0.6 at about $40^{\rm o}<\theta<70^{\rm o}$. For run RHDW0.6, the radial velocity is higher than 10$^4$ km s$^{-1}$ in the $57^{\rm o}<\theta<77^{\rm o}$ angular range and the maximum value is about 2$\times$10$^4$ km s$^{-1}$ at $\sim70^{\rm o}$. For run RMHDW0.6, the radial velocity is higher than 10$^4$ km s$^{-1}$ in the angular $50^{\rm o}<\theta<77^{\rm o}$ range and the maximum value reaches 3.6$\times$10$^4$ km s$^{-1}$ at $\sim63^{\rm o}$. The angular position of the maximum outflow velocity corresponds to the peak of the mass flux density of winds. The high-velocity winds have a column density of $N_{\rm H}>10^{22}$ cm$^{-2}$ and take place on the gray belt, which means that the line force drives out the high-velocity winds.

In run RMHDW0.6a, whose X-ray luminosity is two times that of run RMHDW0.6, the mass outflow rate and power of winds is less than that in run RMHDW0.6, as given in Table 1. In the region of $17^{\rm o}<\theta<62^{\rm o}$, the column density of RMHDW0.6a are less than that of RMHDW0.6, while at other angles their column densities are comparable. This causes the ionization parameter in run RMHDW0.6a to be higher than that in run RMHW0.6 and then the strength of the line force in run RMHDW0.6a is weaker than that in run RMHW0.6. As a result, the velocity of winds in run RMHDW0.6a is less than in run RMHDW0.6. Over $50^{\rm o} <\theta<80^{\rm o}$, the angular-averaged radial velocity at the outer boundary is 0.06 $c$ for RMHDW0.6, while it is 0.05 $c$ for RMHDW0.6a. Observations suggest that the column density and velocity of UFOs are correlated with the X-ray luminosity and hardness (Chartas et al. 2009b; Matzeu et al. 2017). Chartas et al. (2009b) found that flatter X-ray spectra appear to result in lower outflow velocities in the BAL quasar APM 08279+5255, which seems to be explained in the magnetically driven winds model (Fukumura et al. 2010). Matzeu et al. (2017) reported that the outflow velocities increase with the increase of hard X-ray luminosity (7--30 keV) in PDS 456, which seems to be explained well in both radiation-driven models (e.g. King \& Pounds 2003; Matzeu et al. 2017) and a generic magnetic-driving mechanism (Fukumura et al. 2018). In our simulations, the effect of the X-ray spectral shape cannot be studied. Our results show that wind speed does not increase with the increase of X-ray luminosity, which seems to not be consistent with the observation of PDS 456. In line-driving wind models, the increase of the X-ray luminosity is helpful to enhance the ionization parameter and then the strength of line force could be reduced. The inconsistency between our results and the observations of PDS 456 could be due to that when the X-ray luminosity increases, the whole bolometric luminosity could increase. As pointed out by Matzeu et al. (2017), the UFOs' velocity in PDS 456 could be proportional to the whole bolometric luminosity rather than the X-ray luminosity, though these analyses gave the correlations with hard X-ray luminosity, not the bolometric value. It is unfortunate that only XMM-Newton observations provide simultaneous optical/UV/X-ray observations. However, it is not clear whether optical/UV flux is as variable as the observed X-ray flux (Matzeu et al. 2016a; 2016b).

In the model (i.e. run RMHDW0.6h) with higher numerical resolution, the results are qualitatively similar to that in run RMHDW0.6. As shown in Figure 6, the column density in run RMHDW0.6h is almost consistent with that in run RMHDW0.6, while the angle distribution of high-velocity winds is slightly broader in run RMHDW0.6h than in run RMHDW0.6. This demonstrates that the grid resolution in simulations is enough for studying the winds.

\subsection{Role of magnetic field} \label{subsec:magneticfield}
Observations suggest that a large-scale poloidal magnetic field could exist. In radio-loud AGNs and radio galaxies (e.g. M87), the radio jets are measured and are considered to be accelerated and collimated efficiently by a large-scale magnetic field that may originate from either a spinning BH itself or the inner part of the disk. (e.g. Hada et al. 2011; Nakamura et al. 2018; Chen \& Zhang 2021). In radio-quiet Seyfert galaxies, X-ray coronae are often conjectured to be created by local magnetic activities such as magnetic reconnection (e.g. Liu et al. 2002; Kawanaka et al. 2008). Such magnetic activities may well be a part of a global magnetosphere of a BH (e.g. Hirose et al. 2004).

In this paper, we initially adopt a large-scale poloidal magnetic field, whose magnitude is weaker than the gas pressure at the disk surface. Away from the disk surface, the gas density decreases significantly. The magnetic field above the disk surface may become strong at some positions, compared with the gas pressure, because of the significantly decreasing of gas density. Therefore, when winds are formed, the magnetic field is not necessarily weak in the corona. On the other hand, a weakly magnetized Keplerian disk is known to be subject to magneto-rotational instability (MRI; Balbus \& Hawley 1991). This instability can grow fast in the disk. However, the large-scale poloidal magnetic field does not become stronger due to the growth of MRI in the disk because the total magnetic flux across the disk surface is constant. The Keplerian thin disk is treated as a boundary condition of our simulations and the evolution of a thin disk is not considered. Therefore, the presence of MRI in the thin disk does not affect our results much. The formation of the large-scale poloidal magnetic field is an open issue. As adopted in this paper, the large-scale poloidal magnetic field may be weak. The magnetic flux across unit area at some radius can be changed by the inward advection of large-scale flows in the disk and the outward diffusion driven by turbulence in the disk (Guan \& Gammie 2009; Cao \& Spruit 2013; Contopoulos et al. 2017). For simplicity, it is assumed that the outward diffusion of field lines can balance the inward advection of the field lines. Under this assumption, the magnetic flux across the disk surface is also fixed at all radii.

We time averaged the magnetic field over $t=0.5$--1.0 $T_{\rm orb}$. Figure \ref{fig7} shows the two-dimensional structure of the magnetic field for run RMHDW0.6. As shown in panels (A) and (B), in most of the region around the pole and the disk surface, the magnetic pressure is dominated by the poloidal magnetic field and stronger than the gas pressure. In the angular region of $80^{\rm o}<\theta<90^{\rm o}$, the X-ray radiation from the corona is shielded by the winds generated from the disk surface within 40$r_{\rm s}$, which causes the gas outside 40$r_{\rm s}$ to become cool due to the radiation cooling and then the gas pressure becomes weak. In most of the middle-latitude region of $30^{\rm o}<\theta<80^{\rm o}$, the magnetic pressure is dominated by the toroidal magnetic field, while the total pressure is dominated the gas pressure. Panels (C) and (D) show the energy density of the poloidal magnetic field and the magnetic field lines on the $r$--$z$ plane. Compared with the initial magnetic field lines, the field lines at the disk surface become steeper, while the field line away from the disk surface becomes more tilted.

Although the magnetic field is initially set to be weaker than the gas pressure at the disk surface, in the angular range of $80^{\rm o}<\theta<90^{\rm o}$ the magnetic field is stronger than the gas pressure at the large radius. However, because in the region, the line force is much stronger, the Lorentz force is still much weaker than the line force. The Lorentz force, compared with line force, is negligible in directly driving winds. However, in the MHD model, the region around the rotational axis is dominated by magnetic pressure. Especially in the inner region ($<$200 r$_{\rm s}$), the angle range of $\theta<$45$^{\rm o}$ is almost dominated by magnetic pressure. The magnetic pressure prevents gas from spreading to higher latitudes. As shown in Figure 6, the gas column density in the MHD model is lower than that in the HD model on the 0--20$^{\rm o}$ angular range. At middle and low latitudes (20$^{\rm o}$--70$^{\rm o}$), the gas column density in the MHD model is higher than that in the HD model. Higher column density is helpful to obscure the X-ray photons, which causes the line force to be more effective in the MHD model than in the HD model. Therefore, in the MHD model, the effective angular region of the line force is broader than in the HD model, and the magnitude of line force is also stronger than in the HD model.

\subsection{Application to UFOs} \label{subsec:application}
Tombesi et al. (2010) reported that blue-shifted Fe K absorption lines, with high velocities (up to 0.2--0.4 $c$), exist in the X-ray spectra of several radio-quiet AGNs. This implies that the highly ionized absorbing materials are moving fast away from their nuclei. Tombesi et al. (2010) first defined UFOs to describe the highly ionized absorbers. UFOs have outflow velocities higher than 10$^4$ km s$^{-1}$, their ionization parameter is in the range of log($\xi$/(erg s$^{-1}$ cm))$\sim$3--6, and their column density is distributed in $10^{22}$ cm$^{-2}$$\lesssim N_{\rm H}\lesssim$ $10^{24}$ cm$^{-2}$ (Tombesi et al. 2011). More than 35\% of radio-quiet AGNs show UFOs features.

In our simulations, in the HD case, the outflows over the 58$^{\rm o}$--77$^{\rm o}$ angle range satisfy the properties of UFOs. In the MHD case, the angle range becomes 50$^{\rm o}$--77$^{\rm o}$. If the detection probability of UFOs is defined as $\Omega_{\rm UFO}/4\pi$, $\Omega_{\rm UFO}$ is the solid angles of UFOs (Nomura et al. 2016), the detection probability of the HD UFOs is around 35\% and the detection probability of the MHD is around 47\%. This implies that our results are consistent with observations.

Gofford et al. (2015) measured the Fe K absorption of 20 AGNs and found that the mass outflow rates are of the order of $\sim0.01$--1 ${M}_{\odot} {\rm yr}^{-1}$ and the kinetic power is $\sim10^{43{\rm-}45}$ erg s$^{-1}$. As two individual samples, PDS 465 and PG 1211+143 have often been referenced in previous works (e.g. Pounds et al. 2003; 2016; Reevies et al. 2009; 2018; 2020; Danehkar et al. 2018; H\"{a}rer et al. 2021). PDS 456, a radio-quiet quasar with a bolometric luminosity of $L_{\rm bol}\sim 10^{47}$ erg s$^{-1}$, was measured to show the UFOs launched with $\sim$50$r_{\rm s}$ (Reevies et al. 2009). The kinetic power of UFOs is 20\% of the bolometric luminosity, and the mass outflow rate is $\sim 10$ ${M}_{\odot} {\rm yr}^{-1}$ (Nardini et al. 2015). For PG 1211+143, with a bolometric luminosity of $L_{\rm bol}\approx 4\times 10^{45}$ erg s$^{-1}$ and an X-ray luminosity (2--10 keV) of $\sim10^{44}$ erg/s, Pounds et al. (2003) estimated that its UFOs are generated within 130 $r_{\rm s}$ and the mass outflow rate and kinetic energy are $\sim3$ ${M}_{\odot} {\rm yr}^{-1}$ and $\sim5\times10^{44}$ erg s$^{-1}$, respectively. In Gofford et al.'s work (2015), the BH mass of most of samples is comparable with that of our models. Table 1 shows that the mass outflow rates are 0.15--0.26 ${M}_{\odot} {\rm yr}^{-1}$ and the kinetic power is of the order of $\sim10^{43{\rm-}44}$ erg s$^{-1}$. Our results are consistent with Gofford et al.'s results (2015). The BH mass and luminosity of PDS 456 are much greater than that of our models. Our models cannot be used to make comparisons with PDS 456. For PG 1211+143, its luminosity in units of Eddington luminosity is comparable with that of our models. The mass outflow rates in our models are lower than that of PG 1211+143. The physical quantities listed in Table 1 depend on the density at the disk surface. In this paper, we set the density at the disk surface to be $10^{-12}$ g.cm$^{-3}$. When we increased the density at the disk surface, the mass outflow rate and power of the winds increase, and vice versa.

In addition to UFOs being detected at the Fe K bands, they are also detected at soft X-ray bands (Pounds et al. 2016; Serafinelli et al. 2019; Reeves et al. 2020). The soft X-ray UFOs show low column density ($N_{H}\sim10^{20{\rm-}22}$ cm$^{-2}$) and a low-ionization parameter (log($\xi$/(erg s$^{-1}$ cm))$\sim$0--3), which is significantly distinct from those observed in the Fe K UFOs (Serafinelli et al. 2019). The soft X-ray UFOs are considered to be located on parsec scales rather than around the accretion disk (Reeves et al. 2020). The line force becomes significant when $\xi<100$ erg s$^{-1}$. Therefore, if the line force continuously accelerates winds, the winds should always be in the low-ionization state. However, our simulations find that when the low-ionization winds are exposed to X-rays, the low-ionization winds become high-ionization. When the X-rays are shielded by dense gases in the inner region, the degree of ionization of the high-ionization winds becomes so low that line-force driving becomes effective. So this way, the winds are intermittently accelerated by the line force to a high speed. As shown in a snapshot of an ionization parameter (such as panel (D) of figure 4), the high-velocity winds formed around the disk should entrain the low-ionization high-velocity winds. The multiphase high-velocity winds may evolve into a warm absorber on parsec scales, which may produce the blue-shifted lines seen in the soft X-ray bands.

In addition, the Fe K lines in UFOs show a quiet broad width that often requires a phenomenological turbulent motion of $\sim$10000 km s$^{-1}$ or higher (e.g. Nardini et al. 2015; Tombesi et al. 2015; Reeves et al. 2018). The velocity is too high to be explained by local thermal turbulence in the winds. Alternatively, the rotation of the winds could result in broadening of the Fe K line because the projected value of the rotational velocity at the line of sight can be either positive or negative. The rotational velocity depends on the location of winds, while its projected value depends on the cosine value ($\propto cos(\alpha)$) of the angle ($\alpha$) between the sight line and rotational velocity. For run RMHDW0.6, the rotational velocity ($v_{\phi}$) in winds can reach $\sim10000$ km s$^{-1}$ at $\sim420 r_{\rm s}$ while $\sim4500$ km s$^{-1}$ at $\sim1500 r_{\rm s}$. At smaller radii, the rotational velocity becomes higher. Rotation can result in the redshift and blueshift of spectral lines, if the rotation plane is not perpendicular to the sight line. When observations can distinguish the receding side and approaching side of the rotation body, one can detect the redshift lines from the receding side and the blueshift lines from the approaching side. When observations cannot distinguish the receding side and the approaching side, the detected lines are broadened due to rotation, rather than the center of lines being redshifted or blueshifted. The red side of the line center is attributed to the receding side, while the blue side of the line center is attributed to the approaching side. The outflows result in the blueshift of the line center. For the AGNs with high inclination, therefore, the broad Fe K lines could be formed in UFOs because of the fast rotation of UFOs. Another effect, the special relativistic beaming effect, might be important due to the fast rotation of UFOs. The blueshifted components originating from the approaching side of UFOs are preferentially amplified, while the radshifted components from the receding side would be greatly reduced by factor of $1/(\Gamma(1-v_{\phi} cos(\alpha)/c))^3$, where $\Gamma=1/\sqrt{1-v_{\phi}^2/c^2}$. This implies that the redshifted components may well appreciably be suppressed.

On the other hand, Fe K UFOs are observed in broad-line radio galaxies (Tombesi et al. 2014). The UFOs in radio-loud AGNs have similar properties to that of radio-quiet AGNs. Tombesi et al. (2014) reported that the UFOs in radio-loud AGNs are detected at the inclination angle of $\sim10^{\rm o}$ to $\sim70^{\rm o}$ away from the jets. In this paper, our simulations seem to fail to describe such low-inclination UFOs. This may be attributed to the use of a weak magnetic field in our simulations. In the next paper, we will adopt a strong magnetic field. Preliminary results show that when a strong magnetic field is adopted, the high-velocity winds have a larger opening angle than current results. Our next paper is in preparation.

Fe K UFOs are also observed in the faint Seyfert 2 galaxies, such as NGC 2992, whose luminosity is about 0.001--0.04$L_{Edd}$ (Marinucci et al. 2018). Simulations find that when the AGN luminosity is very low, the line-driving mechanism becomes invalid. For example, Nomura et al. (2016) reported that UFOs could exist in any luminous AGNs, whose luminosity is higher than 0.1$L_{\rm Edd}$. According to the results from numerical simulation, the UFOs in NGC 2992 should not be driven by the line force, because the radiation force is too weak to blow away material from the disk surface. Magnetic driving is a promising mechanism to understand the Fe K UFOs of NGC 2992.

\section{Summary and Discussion} \label{sec:Summary}
We performed two-dimensional MHD simulations to study the winds driven by the line force from the thin disk surface. Winds with a velocity higher than 10$^4$ km s$^{-1}$ are found in our simulations. Line force plays an important role in driving high-velocity winds. Compared with the gas pressure at the disk surface, the initial magnetic field is set to be weak. In the computational region, the magnetic field is initially much stronger than the gas pressure, because the initial density is very low. When winds are formed, the ratio of magnetic pressure to gas pressure is significantly changed. In the high-velocity winds, the magnetic pressure is almost weaker than the gas pressure. Compared with the line force, the effect of the Lorentz force is negligible in directly driving winds. However, magnetic pressure dominates the region around the pole and prevents gases from spreading to higher latitudes, which causes the gases to become dense at middle and low latitudes. Although the weak magnetic field is inefficient in directly driving winds, the increase of gas density at middle and low latitudes is helpful to shield the X-ray radiation from the center. This makes the line force more efficient. Therefore, in the MHD model, the line force is more significant and the higher-velocity winds are produced. When the magnetic field becomes strong, it is worth studying the role of magnetic field in driving winds and the properties of winds. In the future, we will consider the case of a strong magnetic field.

In our simulations, dense gases are blown off from the inner disk surface by the line force and Compton-scattering force and the blown-off gases shield the X-ray radiation from the corona. The degree of ionization of the gases in the outer region ($>$100r$_{\rm s}$) is lower than that in the inner region ($<$100r$_{\rm s}$). The magnitude of the line force is weaker in the inner region than in the outer region on the angular range of $\theta>60^{\rm o}$. The X-ray radiation cannot effectively pass through dense gases in the inner region at some angle, and the gas at a large radius becomes weakly ionized. The weakly ionized gases absorb UV photons and then the blown-off gases are further accelerated by the line force. The line force becomes significant on the region like filaments. The location of filaments changes with time and then the winds driven by line force are exposed to the X-rays and become highly ionized. The high-velocity winds are in multiphase, i.e. low-ionization winds (log($\xi$/(erg s$^{-1}$ cm))$\sim$0--3) and high-ionization winds (log($\xi$/(erg s$^{-1}$ cm))$\sim$3--6). In the sense of time average, the high-velocity winds have the properties of UFOs, such as the ionization parameter and column density.

In previous works, Proga et al. (2000), Proga \& Kallman (2004), and Nomura et al. (2016) have implemented HD simulations of AGN winds by the line force. Qualitatively, the properties of the line-force driven winds are quite similar to those found by the mentioned previous works. One significant difference is that the high-velocity winds have a broader opening angle in our simulations than in their simulations. This makes the detection probability of UFOs in our simulations more consistent with observations. The reason for a larger opening angle of winds is as follows. In the present paper, we set the X-ray luminosity according to the relation of the X-ray luminosity and the UV luminosity given by observations by Lusso \& Risaliti (2016). In their papers, the X-ray luminosity is fixed. The X-ray luminosity in this paper is lower than that in their papers. Lower X-ray luminosity can result in a lower ionization parameter in broader regions. Therefore, the region, where the line force can play a role in this paper, is larger than that in their simulations. We further identify the role of a weak large-scale poloidal magnetic field in driving winds, as mentioned above.

In our simulations, the thin disk is considered as the simulation boundary. In fact, there is an interaction between the thin disk and the outflows/winds generated from the thin disk. The outflows can reduce the mass accretion rate through the thin disk, transfer the angular momentum of disk, and then influence the structure of the thin disk. The change in disk structure influences the formation and power of outflows again. However, including the thin disk in global simulations is a challenging task. Ohsuga \& Mineshige (2011) simulated a truncated thin disk, which was composed of a thick disk within $\sim 7 r_{\rm s}$ and a thin disk of 7--25$r_{\rm s}$. Sadowski  (2016) simulated a thin disk within 7$r_{\rm s}$. A global thin disk from 3$r_{\rm s}$ to $\sim$1500$r_{\rm s}$  has not been simulated. When one simulates the winds/outflows, the thin disk is often decoupled from the computational region in previous simulations and taken as the boundary condition. Based on the decoupled simulations, Nomura et al. (2020) used an iterative technique to solve the interaction between outflows and accretion by coupling the mass outflow rate in the computational region and the mass accretion rate in the thin disk. Nomura et al. (2020) found that the wind power is reduced but the wind power is still strong. In this paper, we assume that there is enough material supply at the outer boundary of disk to keep the accretion rate of the thin disk constant. In analytical models, an accretion-ejection model has been suggested in previous works (e.g. Ferreira \& Pelletier 1993; Jacquemin-Ide et al. 2019). Jacquemin-Ide et al. (2019) coupled magnetically driven jets/outflows to a weakly magnetized accretion disk, where MRI structures are considered, and they found a new class of MRI-like driven outflows from the weakly magnetized accretion disk. Such analytical models are worth studying again based on 3D global MHD simulations of turbulent accretion disks with large-scale magnetic fields.

We note that scattered X-ray photons are also contributed to ionizing the gas (Higginbottom et al. 2014). However, the effect of the scattered X-ray photons is neglected in this work and previous works. Higginbottom et al. (2014) pointed out that the effect may be important, and the scattered X-ray photons may make winds so highly ionized that line driving may become not be efficient. However, radiative transfer and hydrodynamics are decoupled in Higginbottom et al. (2014). Real flows could adjust themselves to make scattered photons not to be strong enough to highly ionize winds, so that line driving could be efficient. Simulations of realistic flows need to couple radiative transfer and HD/MHD calculations, which is computationally expensive. Such simulations might need a next generation code of HD/MHD formalism, which can more realistically treat radiative transfer in more detail. Therefore, now the post-process approach is essentially the only plausible way to quantitatively assess the charge state of plasma, as implemented by Higginbottom et al. (2014).

\acknowledgments{This work is supported by the Natural Science Foundation of China (grant 11973018) and Chongqing Natural Science Foundation (grant cstc2019jcyj-msxmX0581). The authors thank the anonymous referee for the constructive suggests.}

\end{document}